\documentclass[aps,pra,twocolumn,showpacs,reprint,superscriptaddress,longbibliography,floatfix]{revtex4-1}

\usepackage{amsmath, amsfonts, amssymb, bm}
\usepackage{mathtools}
\usepackage[inline]{enumitem}
\usepackage{graphicx}
\usepackage{xcolor}
\usepackage[breaklinks,hypertexnames=false]{hyperref}
\hypersetup{colorlinks,linkcolor={magenta},citecolor={blue},urlcolor={blue}} 

\usepackage[capitalize]{cleveref}
\usepackage{multirow}

\usepackage{glossaries}
\glsdisablehyper

\newacronym{nis}{N-I-S}{normal metal--insulator--superconductor}
\newacronym{fis}{F-I-S}{ferromagnet--insulator--superconductor}
\newacronym{nfis}{N-Fi-S}{normal metal--ferromagnetic-insulator--superconductor}
\newacronym{ffis}{F-Fi-S}{ferromagnet--ferromagnetic-insulator--superconductor}

\newacronym{abs}{SABS}{surface Andreev bound states}

\newacronym{bdg}{BdG}{Bogoliubov-de Gennes}

\newacronym{zbp}{ZBCP}{zero-bias conductance peak}

\newcommand{\up}{\uparrow}
\newcommand{\dw}{\downarrow}
\newcommand{\e}{\mathrm{e}}

\newcommand{\si}{\hat{\sigma}_{0}}
\newcommand{\sx}{\hat{\sigma}_{1}}
\newcommand{\sy}{\hat{\sigma}_{2}}
\newcommand{\sz}{\hat{\sigma}_{3}}

\newcommand{\tz}{\hat{\tau}_{3}}

\newcommand{\mbf}[1]{\mathbf{ #1 }}

\DeclarePairedDelimiter\mean{\langle}{\rangle}

\DeclareMathOperator{\sgn}{sgn}

\begin{document}
	
	\title{Thermoelectric detection of Andreev states in unconventional superconductors} 
	
	\newcommand{\aalto}{Department of Applied Physics,
		Aalto University, 00076 Aalto, Finland}
	\newcommand{\uam}{Department of Theoretical Condensed Matter Physics,
		Universidad Aut\'onoma de Madrid, 28049 Madrid, Spain}
	\newcommand{\nagoya}{Department of Applied Physics,
		Nagoya University, Nagoya 464-8603, Japan}
	
	\author{Tony Savander} 
	\affiliation{\aalto}
	
	\author{Shun Tamura}
	\affiliation{\nagoya}
	
	\author{Christian Flindt}
	\affiliation{\aalto}
	
	\author{Yukio Tanaka}
	\affiliation{\nagoya}
	
	\author{Pablo Burset}
	
	\affiliation{\aalto}
	\affiliation{\uam}
	
	\date{\today}
	
	\begin{abstract}
		We theoretically describe a thermoelectric effect that is entirely due to Andreev processes involving the formation of Cooper pairs through the coupling of electrons and holes. The Andreev thermoelectric effect can occur in ballistic ferromagnet-superconductor junctions with a dominant superconducting proximity effect on the ferromagnet, and it is very sensitive to surface states emerging in unconventional superconductors. We consider hybrid junctions in two and three dimensions to demonstrate that the thermoelectric current is always reversed in the presence of low-energy Andreev bound states at the superconductor surface. A microscopic analysis of the proximity-induced pairing reveals that the thermoelectric effect only arises if even and odd-frequency Cooper pairs coexist in mixed singlet and triplet states. Our results are an example of the richness of emergent phenomena in systems that combine magnetism and superconductivity, and they open a pathway for exploring exotic surface states in unconventional superconductors. 
	\end{abstract}
	
	\maketitle
	
	\section{Introduction\label{sec:intro}}
	
	Thermoelectric effects, where temperature gradients and electric voltages are converted into each other, open a promising way to reuse waste heat in electronic devices~\cite{Bauer_2012}. In mesoscopic conductors, a thermoelectric current can be generated by breaking the electron-hole symmetry around the chemical potential. However, the current is usually small at low temperatures. By contrast, large thermoelectric effects have recently been predicted~\cite{Kalenkov_2012,Bergeret_2014,Machon_2014} and measured~\cite{Kolenda_2016,Kolenda_2017} in hybrid junctions between superconductors and magnetic materials. The novel field of superconducting spintronics~\cite{Eschrig_2011,Linder_2015,Eschrig_RPP} explores such phenomena that arise due to the interplay between superconductivity and magnetism~\cite{Keizer_2006,Birge_2010,Robinson_2010,Robinson_2010a,Yang_2010,Hubler_2012}. Interestingly, by coupling the spin degree of freedom to the heat (or charge) transport in ferromagnet-superconductor hybrid junctions, it may be possible to develop more efficient thermoelectric devices~\cite{Giazotto_2014,Bergeret_2015,Giazotto_2018}, such as coolers~\cite{Machon_2014,Bergeret_2014,Sanchez_2018} and thermometers~\cite{Giazotto_2015}.
	
	Many of these applications are based on conventional superconductors with $s$-wave, spin-singlet pair potentials~\cite{Bergeret_RMP2}. By contrast, in unconventional superconductors~\cite{Kashiwaya_RPP,Lofwander_2001}, Cooper pairs couple via anisotropic channels, like $d$-wave~\cite{Hu_1994,Tanaka_1995,Tanaka_1996,Tanaka_1996b}, or form spin-triplet states~\cite{Zwicknagl_1981,Yamashiro_1997,Mackenzie_2003,Maeno_2012,Kallin_2012}. One intriguing aspect of unconventional superconductors is the possibility to develop \gls{abs}, which occur due to the anisotropy of the pair potentials. These edge states can be topologically protected~\cite{Read_2000,Ryu_2002,Sato_2011} and may play the role of Majorana states in condensed matter~\cite{Chiral_SC_2016,Sato_2017,Fujimoto_JPSJ,Aguado_Majo}. \gls{abs} are also related to point or line nodes where the pairing potential vanishes~\cite{Kobayashi_2015,Tamura_2017}, often leading to a pronounced zero-bias peak in the tunneling conductance~\cite{Hu_1994,Shiba_1995,Tanaka_1995,Lofwander_2001}, like the ones observed in, e.g., high-$T_c$ cuprates~\cite{Greene_1997,*Greene_1997err,Alff_1997,Wei_1997,Kashiwaya_1998,Kashiwaya_2000exp,Greene_2002,Hilgenkamp_2008}. 
	
	\begin{figure}[b!]
		\includegraphics[width=0.95\columnwidth]{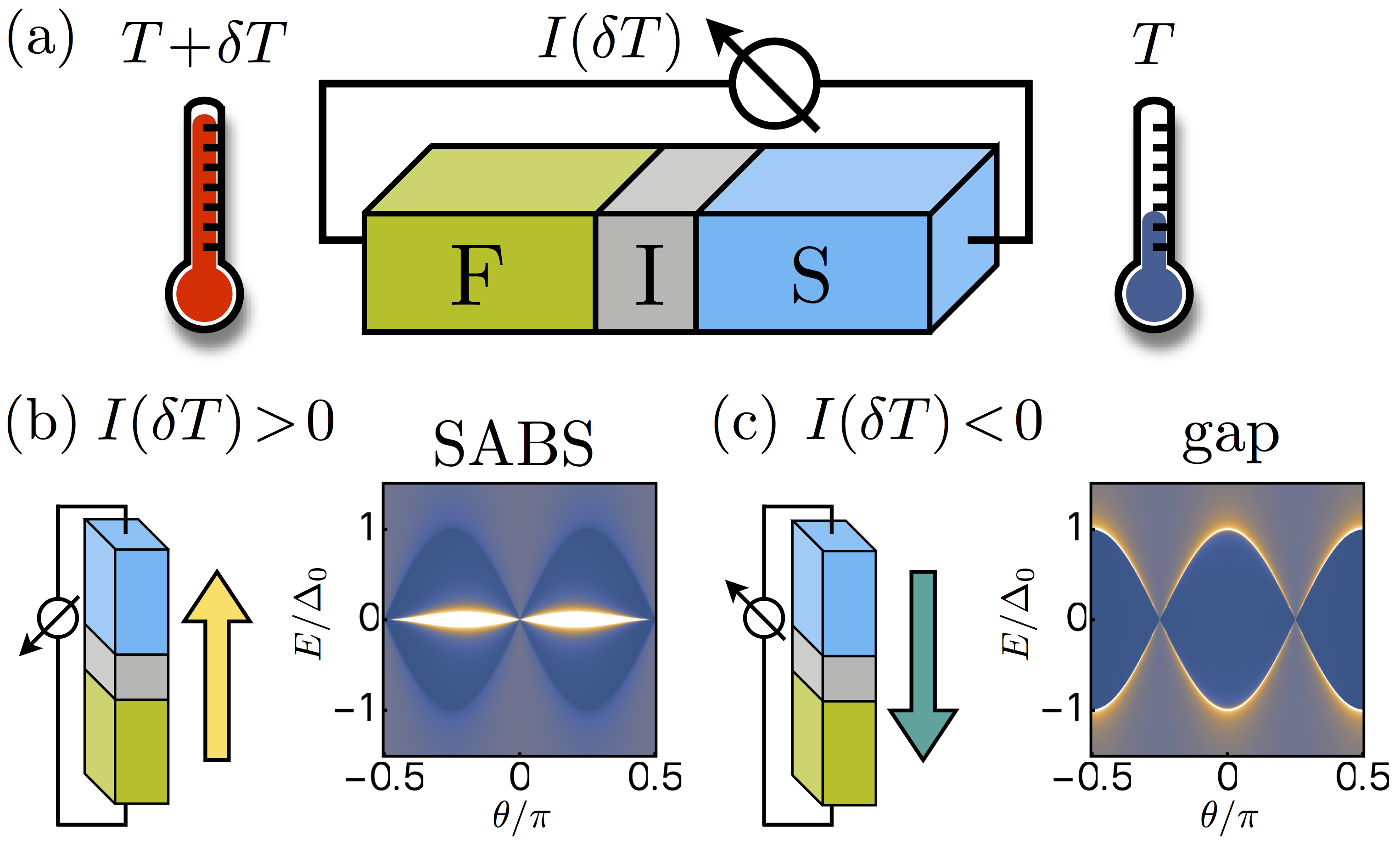}
		\caption{\label{fig:setup} 
			Thermoelectric effect and direction of the current in a ferromagnet-superconductor hybrid junction. 
			(a) Setup indicating the hot and cold reservoirs. 
			(b,c) Assuming a majority of spin-up particles, a positive thermoelectric current flows into the superconductor (b) in the presence of \gls{abs}, while a negative one runs out of it in their absence (c). 
		}
	\end{figure}
	
	Here, we analyze the electrical current induced by a temperature gradient across hybrid junctions involving unconventional superconductors in two and three dimensions, see \cref{fig:setup}. The thermoelectric effect has been measured in ferromagnet-superconductor tunnel junctions under an external magnetic field~\cite{Kolenda_2016} or in proximity to a ferromagnetic insulator~\cite{Kolenda_2017}. For this reason, we concentrate here on the transport properties of ballistic \gls{fis} and \gls{ffis} junctions described by scattering theory~\cite{BTK,Tanaka_1995,Kashiwaya_RPP}. Building on earlier work on the spin and charge transport of \gls{fis} junctions with unconventional pairings~\cite{Sauls_1997,Kashiwaya_1999,Valls_1999,Valls_2000,Yoshida_2000,Tanuma_2002,Tanuma_2002b,Tanaka_2002,Hirai_2003,Tanaka_2009b}, we show that the thermoelectric current is dominated by Andreev processes, where electrons and holes are converted into each other at the surface of the superconductor. Interestingly, the thermoelectric current changes sign in the presence of \gls{abs}: it is positive (it runs into the superconductor) when the conductance features a zero-bias peak and it is negative otherwise, as illustrated in \cref{fig:setup}(b,c). 
	
	Earlier predictions of the thermoelectric effect in ferromagnet-superconductor junctions require that the exchange field from the magnet penetrates the superconductor, and thereby spin-splits its density of states~\cite{Bergeret_RMP2}. Such a magnetic (or inverse) proximity effect can be achieved in thin films of conventional superconductors~\cite{Hubler_2012,Quay_2013}, which could be challenging to implement with unconventional ones. Our results, by contrast, rely on the coupling between the ferromagnet's exchange field and the Cooper pairs leaking out of the superconductor. When the superconducting proximity effect takes place~\cite{Lu_2016,*Lu_2017}, quasiparticles can tunnel into the superconductor only when the pairing rotational symmetry is broken at the interface. However, this additional contribution to the thermoelectric current is always smaller than the Andreev one for pairings featuring zero-bias peaks, and hence it does not affect the direction of the current flow. An analysis of the proximity-induced pair amplitude at the junction interface, based on Green's function techniques, allows us to show that the Andreev thermoelectric current is only possible when even- and odd-frequency Cooper pairs coexist~\cite{Golubov_2007,Eschrig_2007,Golubov_2007b,Golubov_2007c,Tanaka_JPSJ}. The thermoelectric effect is thus a signature of proximity-induced odd-frequency pairing, and it thereby becomes a sensitive tool to explore unconventional superconductivity. 
	
	The rest of the paper is organized as follows. 
	We introduce the theoretical methods in \cref{sec:model} and explain the Andreev thermoelectric effect and the current sign change in \cref{sec:th-effect}. 
	In \cref{sec:odd-w}, we use Green's function techniques to analyze the proximity-induced pairing on the ferromagnetic region. 
	\cref{sec:qp-cont} discusses the quasiparticle contribution to the thermoelectric current. 
	Finally, we present our conclusions in \cref{sec:conc}.

	\section{Theoretical formulation\label{sec:model}}
	
	We consider ballistic junctions between a superconductor and a magnetic normal-state metal separated by an insulating barrier. 
	Transport takes place in the $x$-direction, with the superconductor extending over the region $x\ge 0$. In all cases below, we assume perfectly flat and clean interfaces. 
	Transport in such junctions is described by the \gls{bdg} equations 
	\begin{equation}
	\check{H}\Psi=E\Psi,
	\end{equation}
	with $E$ being the excitation energy. Particle-hole symmetry imposes that every solution $\Psi_E$ of the \gls{bdg} equations with excitation energy $E$ is accompanied by another solution, $\Psi_{-E}$, at energy $-E$. We then choose the basis
	\begin{equation}
	\Psi\!=\![\hat{c}_{\up}(\mbf{k}),\hat{c}_{\dw}(\mbf{k}),\hat{c}^{\dagger}_{\up}(-\mbf{k}),\hat{c}^{\dagger}_{\dw}(-\mbf{k})]^T
	\end{equation}
	in Nambu (particle-hole) and spin space, with $\hat{c}_{\sigma}(\mbf{k})$ being the annihilation operator for an electron with spin $\sigma\!=\up,\dw$ and momentum $\mbf{k}$. The general form of the \gls{bdg} Hamiltonian in any region is
	\begin{equation}\label{eq:hamil}
	\check{H}(\mathbf{k})=\left(\!\begin{array}{cc}
	\hat{h}(\mathbf{k}) \!-\! \mu\si & \hat{\Delta}(\mathbf{k}) \e^{i\phi} \\ \hat{\Delta}^\dagger(\mathbf{k}) \e^{-i\phi} & \mu\si \!-\! \hat{h}^{*}(-\mathbf{k})
	\end{array}\!\right) ,
	\end{equation}
	where $\hat{\sigma}_{0,1,2,3}$ are the usual Pauli matrices in spin space and $\mu$ and $\hat{\Delta}$ the chemical and pair potentials, respectively. It is reasonable to assume that these quantities are constant in each region. For the sake of simplicity, we take the same value of the chemical potential on both sides of the junction. The pair potential is zero in the normal region, which is a good approximation if the superconducting coherence length is much larger than the Fermi wavelength. The superconducting phase $\phi$ only plays a role in junctions with several superconductors, hence, we omit it here. 
	
\begin{figure}
		\includegraphics[width=0.95\columnwidth]{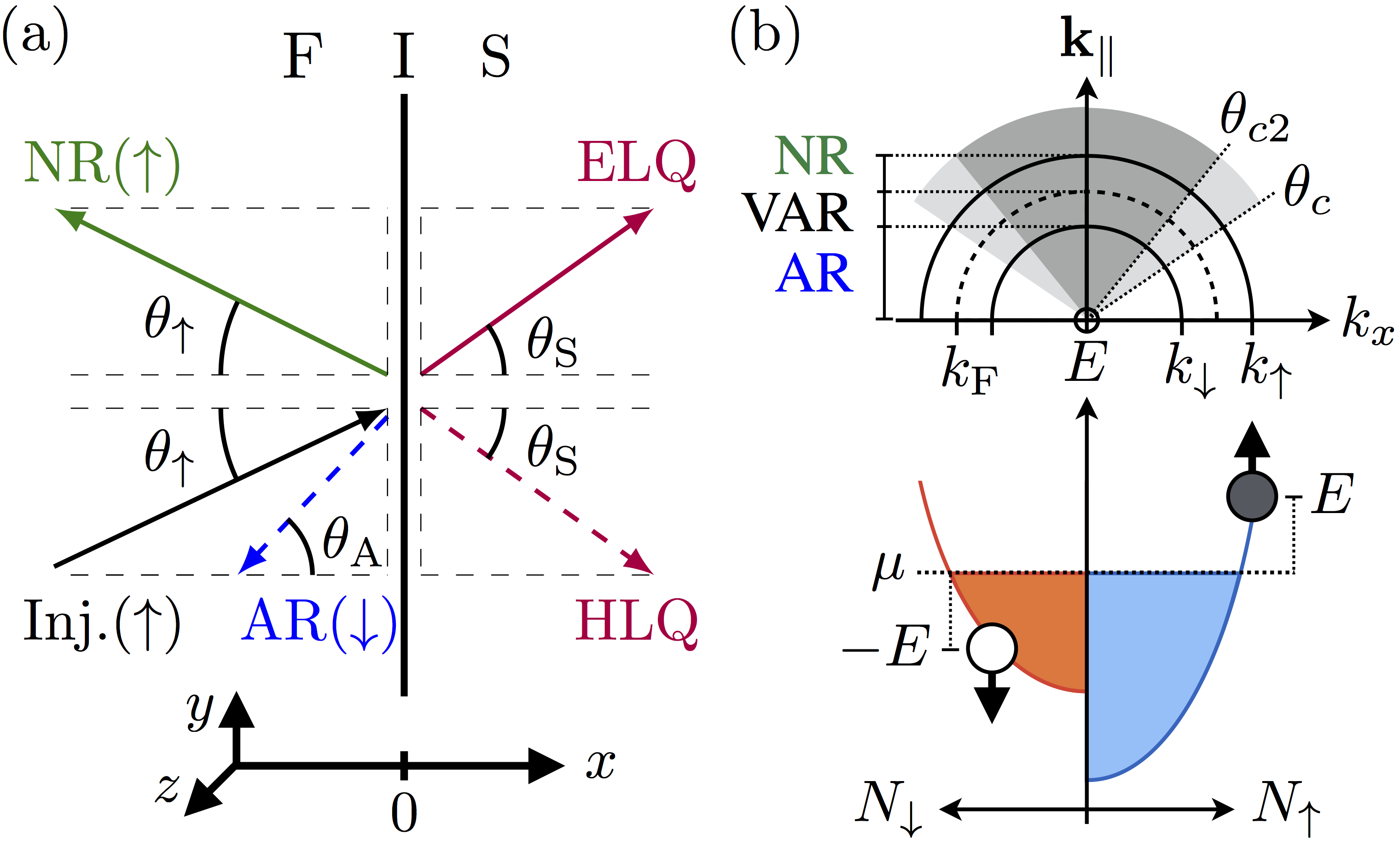}
		\caption{\label{fig:setup2} 
			Andreev reflection in \gls{fis} junctions. 
			(a) An electron incident from the ferromagnet can be normal-reflected (NR), Andreev reflected (AR), or transmitted as an electron- (ELQ) or hole-like (HLQ) quasiparticle. 
			(b) Momentum (top) and energy (bottom) schematics of an Andreev reflection process. Assuming $k_\up>k_\text{F}>k_\dw$, Andreev reflection is only possible for angles of incidence below a critical value $|\theta|<\theta_c$, virtual Andreev reflections (VAR) take place for $\theta_c<|\theta|<\theta_{c2}$, and all incident particles are backscattered for $|\theta|>\theta_{c2}$. 
		}
	\end{figure}

	In the following we consider systems in either two or three dimensions with the single-particle Hamiltonian
	\begin{equation}\label{eq:hamil-sp}
	\hat{h}(\mathbf{r})=  -\frac{\hbar^2 \mbf{\nabla}^2}{2m} \si -U(x)\sz + \left(V_0\si+U_\text{B}\sz\right)\delta(x) ,
	\end{equation}
	where $m$ is the electron mass and $U(x)=U_0\Theta(-x)$ is the exchange potential in the normal-state region with $U_0\ge0$ and $\Theta(x)$ being the Heaviside step function. 
	The potential $V_0$ describes the insulating barrier at the interface with $U_\text{B}$ accounting for the exchange potential of ferromagnetic insulators. Assuming translational invariance in the plane perpendicular to the direction of transport, $\mbf{r}_\parallel=(y,z)$, the momentum parallel to it, $\mbf{k}_\parallel=(k_y,k_z)$, becomes a good quantum number. For a given incident wave, all related wave vectors lie in the same plane, see \cref{fig:setup2}(a). We can then choose the coordinate system so that transport depends only on the azimuthal angle $\theta \!=\! \sin^{-1}(|\mbf{k}_\parallel|/k_\text{F})$, where $k_\text{F}=\sqrt{2m\mu}/\hbar$ is the Fermi wave vector. 
	
	Inserting \cref{eq:hamil-sp} into \cref{eq:hamil}, the \gls{bdg} equations can be decoupled into two spin channels if the pair potential $\hat{\Delta}$ is either a diagonal or an off-diagonal matrix, i.e., proportional to $\hat{\sigma}_{0,3}$ or $\hat{\sigma}_{1,2}$, respectively. The spin-singlet state, $\hat{\Delta}(\mbf{k})\!=\!\Delta_0(\mbf{k})(i\sy)$ always fulfills this condition. The triplet state, which we parametrize as $\hat{\Delta}(\mbf{k})\!=\!\mbf{d}(\mbf{k}) \cdot\mbf{\hat{\sigma}}(i\sy)$, with $\mbf{\hat{\sigma}}$ being a vector of Pauli matrices, fulfills the condition if $\mbf{d}$ lies either along the $z$-axis or in the $x$-$y$ plane. In the following, we only consider pair potentials that fulfill these conditions and that couple electrons with opposite spins. Therefore, the injected and Andreev reflected particles have opposite spins and, if $U_0\neq 0$, Andreev retro-reflection does not take place~\cite{Kashiwaya_1999}. 
	Following a quasiclassical approximation, where the magnitude of the wave vectors is evaluated at the Fermi surface, translational symmetry at the interface entails that
	\begin{equation}\label{eq:Snell-law}
	k_\sigma\sin\theta=k_\text{F}\sin\theta_\text{S}=k_{\bar{\sigma}}\sin\theta_\text{A}^\sigma ,
	\end{equation}
	where $k_\sigma=k_\text{F}\sqrt{1+\rho X}$, with $X=U_0/\mu$ being the polarization from the exchange field and $\rho=1\,(-1)$ and $\bar{\sigma}=\dw(\up)$, when $\sigma=\up(\dw)$. Consequently, assuming up-spin injection and $X>0$, total (normal) reflection takes place for $\theta>\sin^{-1}(k_\text{F}/k_\up)\equiv\theta_{\text{c}2}$, when $k_\text{F}\!<\!k_\up$, and the Andreev-reflected wave becomes evanescent for the angles $\theta_{\text{c}2}>\theta>\sin^{-1}(k_\dw/k_\up)\equiv\theta_{\text{c}}$, if $k_\dw<k_\text{F}<k_\up$. 
	The latter case corresponds to virtual Andreev reflections~\cite{Kashiwaya_1999}, which do not contribute to the current, see \cref{fig:setup2}. 
	
	For a given spin channel $\sigma$, the general solution for the decoupled \gls{bdg} equations is given by
	\begin{equation}
	\Psi_\sigma(x,\mbf{r}_\parallel)= \psi_\sigma(x) \e^{i \mbf{k}_\parallel \cdot \mbf{r}_\parallel},
	\end{equation} 
	where we have defined
	\begin{equation}
	\begin{split}
	\psi_\sigma(x)= \sum\limits_{\alpha=\pm} &\left[ 
	m^\alpha_\sigma \begin{pmatrix} 1 \\ (\eta^\alpha_\sigma)^* \Gamma_\sigma^\alpha \end{pmatrix} \e^{i k_{e,\sigma}^\alpha  \cos\theta_\alpha x} 
	\right.  \\ & \left. +
	\tilde{m}^\alpha_\sigma \begin{pmatrix} \eta^\alpha_\sigma \Gamma_\sigma^\alpha \\ 1 \end{pmatrix} \e^{i k_{h,\sigma}^\alpha \cos\theta'_\alpha x}
	\right] ,
	\label{eq:gen-sol}
	\end{split}
	\end{equation}
	with $m^\alpha_\sigma$ ($\tilde{m}^\alpha_\sigma$) being the amplitude for an electron-like (hole-like) quasiparticle with direction of propagation $\alpha=\pm$. Here, $\theta_+\!=\!\theta$ and $\theta_-=\pi-\theta$ label right and left movers, respectively, and we have defined $\eta_\sigma^\alpha\!=\!\Delta^\alpha_\sigma/|\Delta^\alpha_\sigma|$, $\Gamma_\sigma^\alpha\!=\!|\Delta^\alpha_\sigma|/(E+\Omega_\sigma^\alpha)$, and
	\begin{equation}
	\Omega_\sigma^\alpha = \left\{ \begin{array}{cr}
	\sgn(E)\sqrt{E^2-|\Delta^\alpha_\sigma|^2} & , |E|>|\Delta^\alpha_\sigma| \\
	i\sqrt{|\Delta^\alpha_\sigma|^2-E^2} & , |E|\le|\Delta^\alpha_\sigma|
	\end{array}\right. ,
	\end{equation} 
	where $k_{\text{e(h)},\sigma}^\alpha =k_\text{F}\sqrt{1+(-)\Omega_\sigma^\alpha/\mu+(-)\rho X}$ and $\Delta^\alpha_\sigma=\Delta_\sigma(\theta_\alpha)$ is the pair potential felt by an $\alpha$-moving particle in spin channel $\sigma$. 
	
	Figure \ref{fig:setup2}(a) shows the scattering of an electron incident from the normal-state region into an Andreev reflected hole (AR), a normal reflected electron (NR), an electron-like quasiparticle transmitted to the superconductor (ELQ), or a hole-like transmitted quasiparticle (HLQ). 
	The reflection amplitudes are obtained evaluating the wave function in \cref{eq:gen-sol} for this scattering process and imposing the boundary conditions
	\begin{subequations}\label{eq:bbcc}
		\begin{align}
		\psi_\sigma(0^-) ={}& \psi_\sigma(0^+) , \\
		k_\text{F}\hat{Z}\psi_\sigma(0^-) ={}&  \left.\partial_x \psi_\sigma\right|_{x=0^+} -  \left.\partial_x\psi_\sigma\right|_{x=0^-} ,
		\end{align}
	\end{subequations}
	with 
	\begin{equation}
	\hat{Z}=Z_0\si+Z_m\sz=2m(V_0\si+U_\text{B}\sz)/(\hbar^2 k_\text{F}).
	\end{equation}
	When solving \cref{eq:bbcc}, we follow the Andreev approximation, discarding terms of order $E/\mu$ and $\Delta_0/\mu$, resulting in $k_{\text{e(h)},\sigma}^\alpha \approx k_\text{F}$ in the superconducting region ($x>0$) and $k_{\text{e(h)},\sigma}^\alpha \approx k_{\sigma(\bar{\sigma})}$ for $x<0$. Accordingly, for the scattering of an incident electron with spin $\sigma$ the angles in \cref{eq:gen-sol} become $\theta=\theta'=\theta_{\text{S}}$, for $x>0$, and $\theta$ and $\theta'=\theta^\sigma_{\text{A}}$ for $x<0$. 
	
	The angle-resolved conductance per spin reads~\cite{BTK,Kashiwaya_1999}
	\begin{equation}\label{eq:cond-btk}
	\sigma_\sigma(E,\mbf{k}_\parallel)= 1 - |b_\sigma|^2 + \lambda_\sigma |a_\sigma|^2 \equiv 1-B_\sigma+A_\sigma, 
	\end{equation}
	where $a_\sigma(E,\mbf{k}_\parallel)$ and $b_\sigma(E,\mbf{k}_\parallel)$ are the Andreev and normal reflection amplitudes, respectively, and we have defined $\lambda_\sigma\!=\!\mathrm{Re} \{ k_{\bar{\sigma}} \cos\theta_\text{A}^\sigma/(k_\sigma\cos\theta) \}$ and $\theta_\text{A}^\sigma\!=\!\sin^{-1} ( k_\sigma\sin\theta / k_{\bar{\sigma}} )$. 
	By averaging over all incident angles, the normalized conductance becomes
	\begin{equation}\label{eq:cond-tot}
	\sigma(E) = R_\text{N} \mean{P_\up \sigma_\up(E,\mbf{k}_\parallel) + P_\dw \sigma_\dw(E,\mbf{k}_\parallel)}_{\mbf{k}_\parallel}, 
	\end{equation}
	and the normal state transmissivity is
	\begin{equation}\label{eq:cond-normal}
	\mean{\sigma_\text{N}}_{\mbf{k}_\parallel} = R^{-1}_\text{N}= \left\langle \sum\limits_{\sigma=\up,\dw} \frac{4 P_\sigma \lambda_\text{N}^\sigma }{\left| 1+\lambda_\text{N}^\sigma + \frac{Z_0+\rho Z_m}{\cos\theta_{\text{S}}} \right|^2} \right\rangle_{\mbf{k}_\parallel} , 
	\end{equation}
	with $\lambda_\text{N}^\sigma\!=\! k_\sigma\cos\theta/(k_\text{F}\cos\theta_{\text{S}})$ and $P_\sigma\!=\!(1+\rho X)/2$. 
	The angle averages are given by~\cite{Yamashiro_1998}
	\begin{subequations}\label{eq:ang-avg}
		\begin{align}
		\mean{X_\sigma(\theta)}_{\mbf{k}_\parallel} ={}& \frac{k_\sigma}{2k_\text{F}} \!\!\int\limits^{\pi/2}_{-\pi/2} \!\!\! d\theta\cos\theta X_\sigma(\theta) , \\
		\mean{X_\sigma(\theta,\phi)}_{\mbf{k}_\parallel} ={}& \frac{k_\sigma^2}{\pi k_\text{F}^2} \int\limits^{2\pi}_{0} \! d\phi \!\!\int\limits^{\pi/2}_{0} \!\!\! d\theta\sin\theta \cos\theta X_\sigma(\theta,\phi) , 
		\end{align}
	\end{subequations}
	for two and three dimensions, respectively. 
	Finally, the electrical current is 
	\begin{align}\label{eq:ch-current}
	I\left(\delta T,V\right)={}& \frac{1}{eR_\text{N}} \int_{-\infty}^{\infty}\! dE\, \delta f(E,\delta T,V) \sigma(E) \notag \\
	={}& \sum\limits_{\sigma=\up,\dw} \frac{P_\sigma}{eR_\text{N}} \int_{-\infty}^{\infty}\! dE\, \delta f \mean{2A_\sigma + T^\text{qp}_{\sigma}}_{\mbf{k}_\parallel} \notag \\
	={}& 2I_\text{A}\left(\delta T,V\right) + I_\text{qp}\left(\delta T,V\right) ,
	\end{align}
	where $e$ is the electron charge and $\delta f(E,\delta T,V)\!=\!f(E-eV,T+\delta T) - f(E,T)$ is the difference of the Fermi distributions of the normal-state and superconducting contacts. Here, $f(E,T)\!=\![1+\exp(E/k_\text{B}T)]^{-1}$ is the Fermi function at temperature $T$, $V$ is the applied voltage, and $k_\text{B}$ is Boltzmann's constant. 
	We have used the unitarity of the scattering matrix to define the quasiparticle transmission probability into the superconductor as $ T^\text{qp}_{\sigma} \!=\! 1- B_\sigma - A_\sigma$. Consequently, in \cref{eq:ch-current} we can identify the two main contributions to the current coming from Andreev processes ($2I_\text{A}$) and quasiparticle transfers ($I_\text{qp}$), respectively. 
	Realistic hybrid junctions, built from different materials, will always feature some backscattering at the interface~\cite{Klapwijk_2014}, which affects both $A_\sigma$ and $T^\text{qp}_\sigma$. Therefore, in the following, we consider a finite barrier $Z_0\neq0$, unless otherwise specified. 
	
	If the junction is subjected to a temperature gradient $\delta T$ in the absence of an applied voltage, $V=0$, the difference of the Fermi distributions in each contact is an odd function of the energy $E$, i.e.,
	$\delta f(E,\delta T,0)\!=\!-\delta f(-E,\delta T,0)$. 
	The thermoelectric current is then the integral of \cref{eq:cond-tot} multiplied by an odd function of the energy. Therefore, only the part of \cref{eq:cond-tot} that is odd in the energy will contribute to the integral. 
	Using the anti-symmetry of the difference of Fermi functions, we then find
	\begin{equation}\label{eq:ch-current_sym}
	e R_\text{N} I\left(\delta T,0\right)= \int_{0}^{\infty} dE\, \delta f\left(E,\delta T,0\right) \left[\sigma(E)-\sigma(-E) \right],
	\end{equation}
	which vanishes if $\sigma(E)=\sigma(-E)$, as in the case of conventional \gls{nis} junctions. On the other hand, a thermoelectric current can be induced by an asymmetry in the conductance, as we explain in the following section. 
	
	\begin{figure}
		\includegraphics[width=0.95\columnwidth]{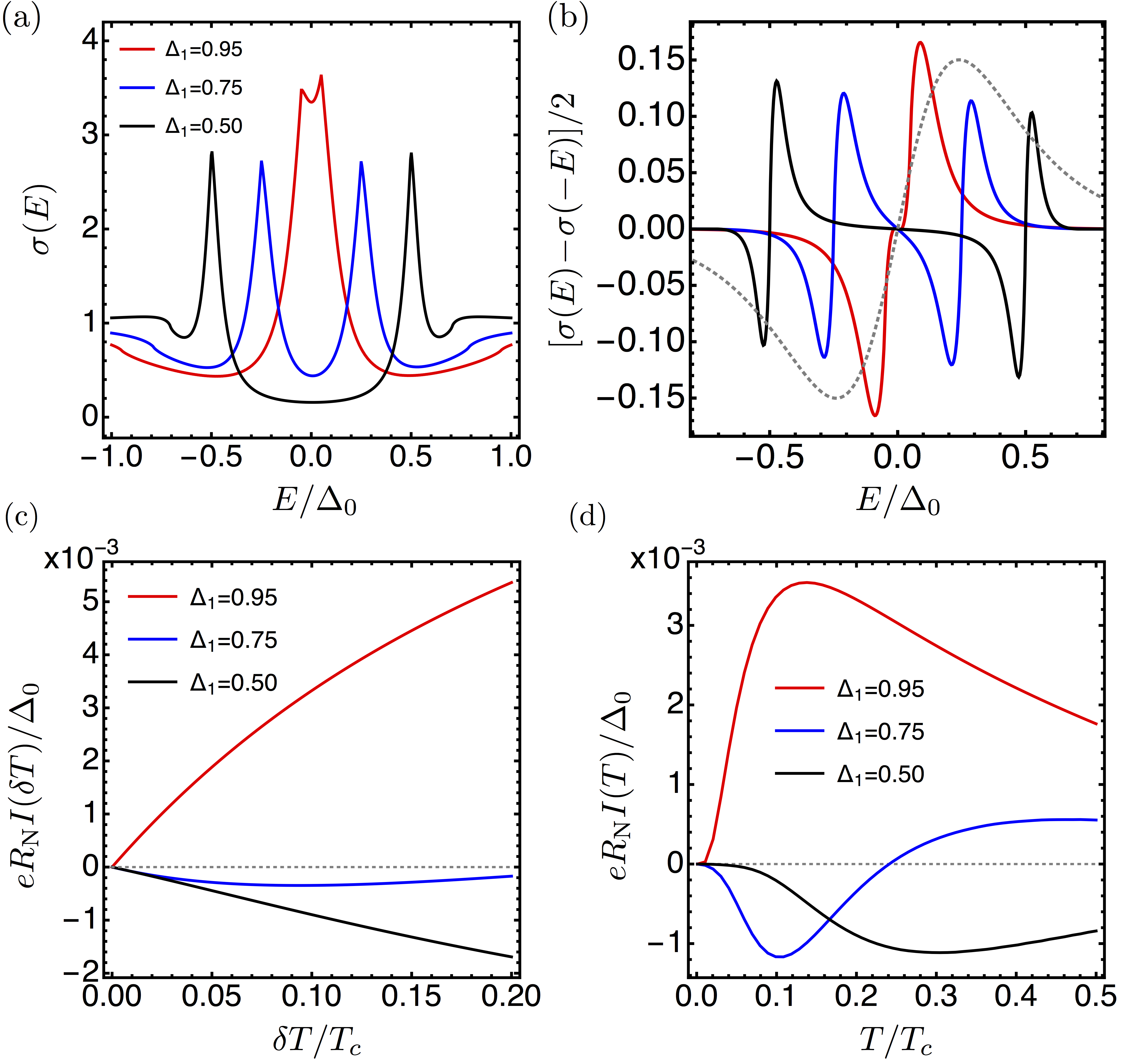}
		\caption{\label{fig:2dpair_test} 
			Conductance (a), its antisymmetric part (b), and the thermoelectric current (c,d) for a two-dimensional junction with $d_{xy}+is$-wave pairing. 
			The dotted line in (b) is the difference in Fermi distribution functions for $T\!=\!0.2T_\text{c}$ and $\delta T\!=\!0.2T_\text{c}$. 
			(c,d) Thermoelectric current as a function of the temperature bias for $T\!=\!0.2T_c$ (c) and the base temperature with $\delta T\!=\!T/2$ (d). 
			In all cases, we have used $X=0.5$, $Z_0=3$ and $Z_m=0$. 
		}
	\end{figure}
	
	\section{Andreev thermoelectric effect\label{sec:th-effect}}
	
	To illustrate the effect of \gls{abs} on the thermoelectric current we start by considering a singlet $(d_{xy}+is)$-wave pairing,
	\begin{equation}\label{eq:pairing-mix-dis}
	\hat{\Delta}_{d+is}(\theta) = \Delta_\text{T} \left[ \Delta_1\sin(2\theta) +i \Delta_2 \right] i \sy ,
	\end{equation}
	with $\Delta_1+\Delta_2= 1$. Henceforth, the pairing temperature dependence is included in the amplitude by the interpolation formula $\Delta_\text{T}/\Delta_0 \!=\! \tanh(1.74 \sqrt{T_{c}/T-1})$. 
	This pair potential was proposed as a fragile state~\cite{Kashiwaya_2004,Hilgenkamp_2008} that could explain the splitting of the zero-bias peak in high-$T_c$ superconductors~\cite{Greene_2002}. 
	We choose it because the parameter $\Delta_1\!=\!1-\Delta_2$ controls the evolution from a gap profile ($s$-wave, $\Delta_1\!=\!0$) into a zero-energy state ($d_{xy}$-wave, $\Delta_1\!=\!1$). 
	The tunnel spectroscopy of the $(d_{xy}+is)$-wave state, \cref{fig:2dpair_test}(a), reveals that the zero-energy peak in the conductance from the $d_{xy}$-wave pairing splits in the presence of a small $s$-wave component until the Andreev resonances reach the gap edges at $E=\pm\Delta_0$, when $\Delta_1=0$. 
	
	The splitting of the zero-bias peak becomes asymmetric with the energy for an \gls{fis} junction with $X\!\neq\!0$. The energy-antisymmetric part of the resulting conductance, $[\sigma(E)-\sigma(-E)]/2$ in \cref{fig:2dpair_test}(b), presents a Fano-like sign change at the resonance positions. 
	While well-separated resonances feature two Fano-like line shapes, one for positive and another for negative energies, there is only one Fano-like resonance when the \gls{abs} merge around zero energy ($\Delta_1\!\lesssim\!1$). 
	This different behavior for \textit{separated} and \textit{merged} resonances explains the sign change in the current, \cref{fig:2dpair_test}(c,d). 
	For a merged resonance, the anti-symmetric contribution to the conductance around zero energy follows the same profile as the difference in Fermi distributions $\delta f$ [indicated by a gray dotted line in \cref{fig:2dpair_test}(b)]. 
	Consequently, the thermoelectric current is positive after integrating over the energy. 
	By contrast, for separated resonances located at $E=\pm \delta E$, with $\delta E>0$, the anti-symmetric part of the conductance has the opposite sign to $\delta f$ in the energy range $|E|<\delta E$. 
	For experimentally relevant temperatures~\cite{Kolenda_2016,Kolenda_2017}, $T\lesssim T_c/2$ and $\delta T\lesssim0.3T_c$, this energy window provides the biggest contribution to the integral, and the resulting current is thus negative, see \cref{fig:2dpair_test}(c,d). 
	The transition between these two regimes depends on the energy splitting between the \gls{abs}, with $\delta E\gtrsim\Delta_0/4$ enough to avoid a temperature-induced sign change. 
	
	We stress here that the asymmetry in the conductance only occurs at subgap energies, where the normal and Andreev probabilities for each spin channel, $B_\sigma=|b_\sigma|^2$ and $A_\sigma=\lambda_\sigma|a_\sigma|^2$, respectively, cf. \cref{eq:cond-btk}, fulfill $B_\sigma=1-A_\sigma$. 
	Consequently, the conductance is only due to Andreev processes, i.e., $\sigma= 2R_\text{N}\mean{P_\up A_\up+P_\dw A_\dw}_{\mbf{k}_\parallel}$, and the resulting thermoelectric current in \cref{eq:ch-current} is Andreev-dominated, that is,
	\begin{equation}\label{eq:current-andreev}
	I(\delta T,V=0) = 2I_\text{A}(\delta T,V=0) .
	\end{equation}
	We have thus demonstrated the Andreev thermoelectric effect for an \gls{fis} junction with $X\neq0$ and $(d_{xy}+is)$-wave pairing and how the current goes from positive to negative in the presence or absence, respectively, of low-energy \gls{abs}. 
	Within our approach, the necessary and sufficient condition for the Andreev thermoelectric effect is to break time-reversal and particle-hole symmetries per spin species so that the tunnel conductance becomes asymmetric with respect to the energy, cf.~\cref{eq:ch-current_sym}. This can be achieved for any pairing state in an \gls{fis} junction with $X\neq0$, as a result of the different polarization of each spin channel shown in \cref{eq:cond-tot}. 
	
	\begin{table}
		\begin{tabular}{ c c c c }
			\hline\hline
			Spin & wave  & $\hat{\Delta}(\theta_\pm)$ & SABS 
			\\
			\hline\noalign{\smallskip}
			Singlet &\parbox[c]{1.3cm}{$s$} & $\Delta_\text{T}i\sy$ & $\times$
			\\
			Triplet & \parbox[c]{1.3cm}{$p$} & $(\pm\Delta_1\cos\theta+i\Delta_2\sin\theta)\Delta_\text{T}\sx$ & $\Delta_1\neq0$
			\\ 
			Singlet & \parbox[c]{1.3cm}{$d$} & $(\Delta_1\cos2\theta\pm i\Delta_2\sin2\theta)\Delta_\text{T}i\sy$ & $\Delta_2\neq0$
			\\ 
			\hline
		\end{tabular}
		\caption{\label{tab:2dpair_param} 
			Pairing potentials indicating the spin state, angular dependence and the condition for the emergence of \gls{abs}. 
		}
	\end{table}
	
	\begin{figure*}
		\includegraphics[width=0.90\textwidth]{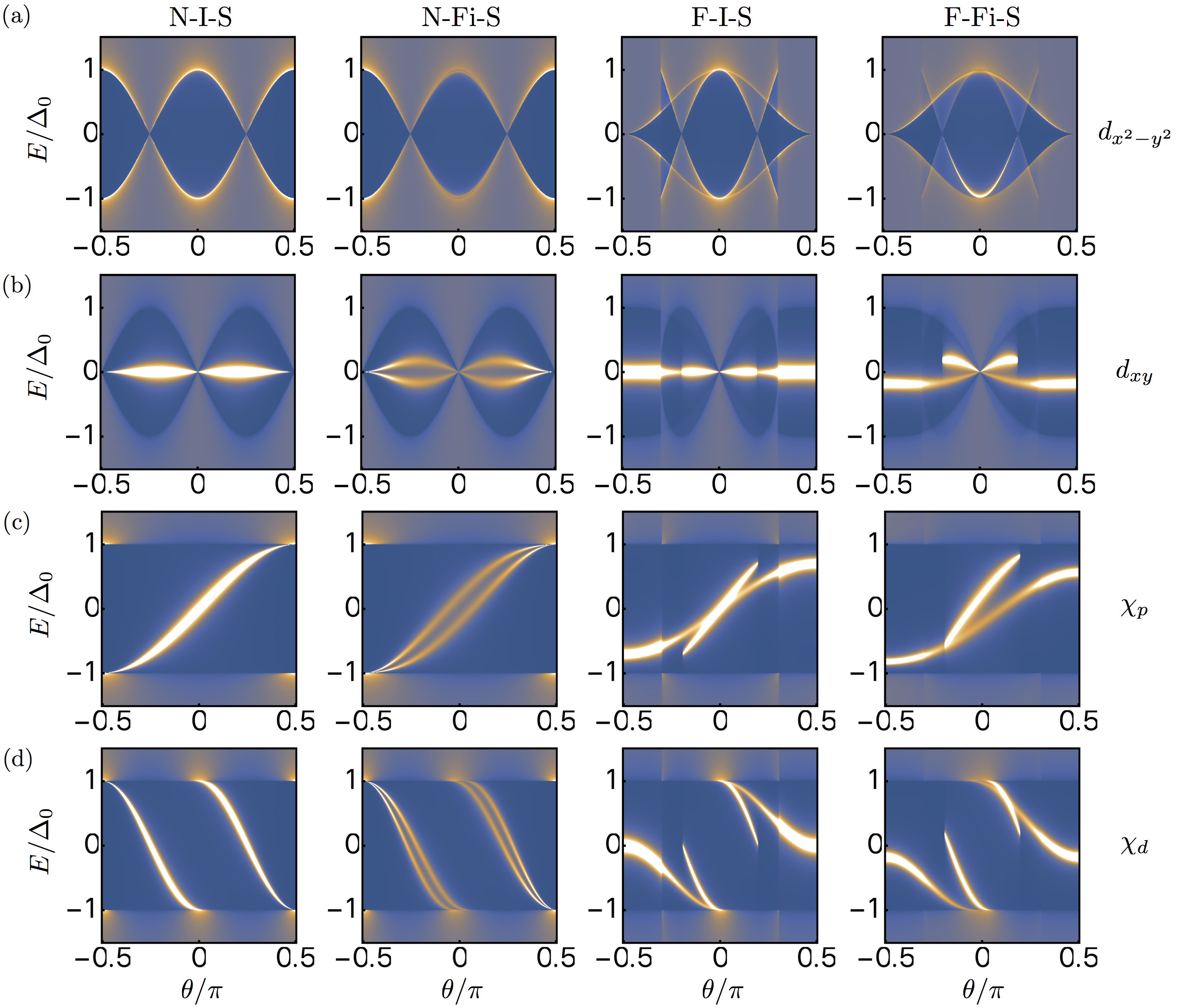}
		\caption{\label{fig:2dpair_map} 
			Spectral differential conductance $(\sum_{\sigma}P_\sigma\sigma_\sigma/\sigma_\text{N})$, where bright colors represent large conductances, for different junctions with the two-dimensional pair potentials (a) $d_{x^2-y^2}$-wave, (b) $d_{xy}$-wave (c) chiral $p$-wave ($\chi_p$) and (d) chiral $d$-wave ($\chi_d$). \gls{nis} and \gls{nfis} (\gls{fis} and \gls{ffis}) junctions have $X\!=\!0$ ($X\!=\!0.5$) and $Z_m=0$ and $Z_m=2.5$, respectively. 
			We set $Z_0\!=\!5$ in all cases. 
		}
	\end{figure*}
	
	Indeed, the Andreev thermoelectric effect can occur in well-known unconventional superconductors. The $(d+is)$-wave pairing of our example explains a specific state of high-$T_c$ superconductors, which, more generally, feature an anisotropic $d$-wave singlet pair potential~\cite{Kashiwaya_RPP,[{For simplicity, we do not consider here mixed singlet-triplet states that could arise from magnetic fluctuations, }] Kontani_2020}. 
	Spin-triplet $p$-wave states have been suggested as an explanation of the properties of UPt$_3$, Sr$_2$RuO$_4$ and other Ru-based compounds~\cite{Mackenzie_2003,Maeno_2012,Kallin_2012}. 
	Some of these materials can develop topological phases, with the emergence of chiral states~\cite{Kallin_2012}. 
	Most of these cases can be described by the pairing potentials listed in \cref{tab:2dpair_param}. Some three-dimensional (3D) chiral superconductors feature more complicated potentials that we detail later. In this section, we assume that the pairing rotational symmetry is preserved at the junction interface. We also define the effective pairing potentials for the right-moving electron- and hole-like quasiparticles in the superconducting region as $\Delta_\sigma^+$ and $\Delta_\sigma^-$, respectively, see \cref{fig:setup2}(a). Rotational symmetry guarantees that $|\Delta_\sigma^+|=|\Delta_\sigma^-|$, and transmitted quasiparticles with the same spin thus experience the same pairing. 
	
	The spectral differential conductance, \cref{fig:2dpair_map}, illustrates the effect of the exchange fields on the subgap Andreev states for the four types of junctions considered here. 
	For a tunnel \gls{nis} junction, it reveals the dispersion of the superconducting pairing. For example, $d_{x^2-y^2}$-wave features a nodal gap without \gls{abs}, while $d_{xy}$-wave presents the same anisotropy but displays zero-energy states, cf.~\cref{fig:2dpair_map}(a,b). The chiral $p$- and $d$-wave states, \cref{fig:2dpair_map}(c) and \cref{fig:2dpair_map}(d), respectively, have \gls{abs} with linear dispersions, but only the chiral $p$-wave state crosses the zero energy point at zero momentum. 
	In all cases, a ferromagnetic insulating barrier (\gls{nfis} junction) causes a spin-splitting of the spectrum, very weak for the gapped $d_{x^2-y^2}$-wave case, but pronounced if \gls{abs} are present. 
	The splitting, however, does not cause any asymmetry. Consequently, there is no current flowing from a temperature bias in an \gls{nfis} junction. 
	
\begin{figure}
		\includegraphics[width=0.95\columnwidth]{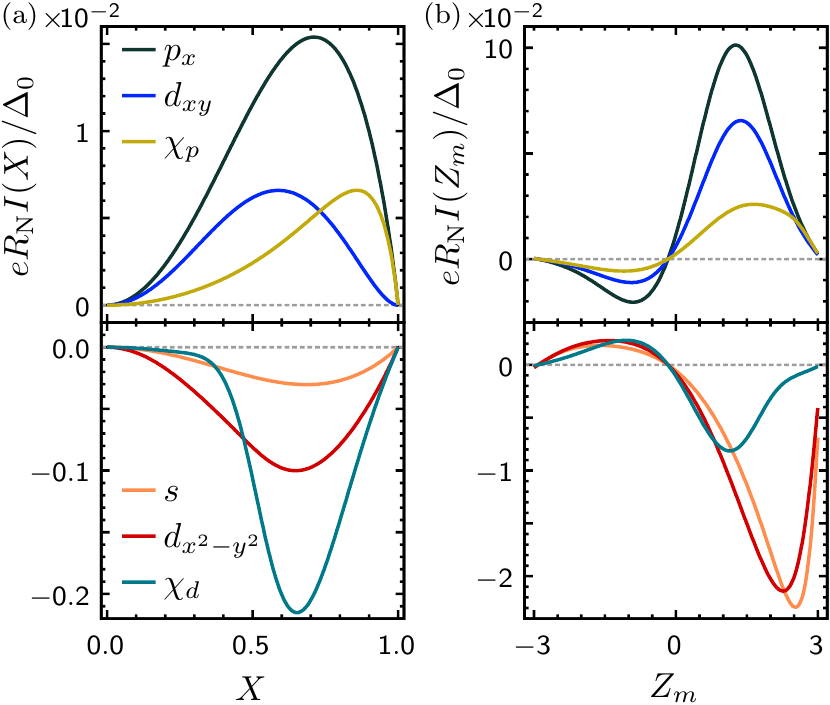}
		\caption{\label{fig:2dpair_param} 
			Thermoelectric effect for the two-dimensional pair potentials listed in \cref{tab:2dpair_param}. 
			(a) \gls{fis} junction with $Z_m=0$. 
			(b) \gls{ffis} junction with $X=0.5$. 
			In all cases, $Z_0=3$, $T=0.2T_\text{c}$, and $\delta T=0.2T_\text{c}$. 
		}
	\end{figure}

	A spin-active barrier alone is not enough to generate a thermoelectric effect; breaking time-reversal and particle-hole symmetries per spin species requires at least an \gls{fis} junction. 
	When $X\!\neq\!0$, the angle of incidence $\theta$ for each spin channel becomes different, see \cref{fig:setup2}(b). The maps in \cref{fig:2dpair_map} show the superposed contribution of both channels, \cref{eq:cond-btk}, revealing a strong angular asymmetry. 
	In \gls{ffis} junctions, the spin-active barrier further splits the spin bands, which, in addition to the angular asymmetry, greatly affects the thermoelectric current. 
	If the spin-active barrier favors the majority spin species, in our case when $Z_m\!>\!0$ if $X\!>\!0$, the asymmetry at subgap energies is enhanced for all pairings, see \cref{fig:2dpair_map}. The resulting current is enhanced by an order of magnitude. 
	Otherwise, when $Z_m\!<\!0$ for $X\!>\!0$, the current flow is reversed, see \cref{fig:2dpair_param}. The resulting thermoelectric currents still have opposite signs for gapped pairings and those featuring \gls{abs}, but their magnitude is smaller than for the case with parallel exchange fields. Notice that the crossover point takes place at $Z_m<0$, since there is a finite current at $Z_m=0$, cf. \cref{fig:2dpair_param}(a). 
	
	\cref{fig:2dpair_param} shows how the sign of the thermoelectric current is positive (negative) for superconducting pairings featuring low-energy \gls{abs} (gap) in the conductance. 
	To explain the direction of the current flow for the chiral $p$- or $d$-wave pairings, we need to take into account that, for a given angle $\theta$, the \gls{abs} can be located at any subgap energy, see \cref{fig:2dpair_map}(c,d). After the angle-average, however, the biggest contributions come from the angles around normal incidence ($\theta\!\sim\!0$). Therefore, the chiral $p$-wave pairing follows the behavior of a merged resonance, with a positive current, while the chiral $d$-wave case behaves like a separated one and thus provides a negative current. 
	This different behavior of the chiral states is not always captured by the tunneling conductance, especially for 3d chiral superconductors~\cite{Tamura_2017}. 
	
	The sign of the Andreev thermoelectric current thus pinpoints the presence or absence of low-energy Andreev states. 
	If the \gls{abs} are exactly at zero energy, like for $p_x$- and $d_{xy}$-wave cases, the positive current is usually one or two orders of magnitude stronger than the negative one for gapped pairings, see \cref{fig:2dpair_map}. 
	We note here that the negative current for gapped superconductors is already much larger than the expected thermoelectric current for equivalent, non-superconducting junctions~\cite{Kalenkov_2012,Bergeret_2014,Machon_2014}\footnote{The magnitude of the non-superconducting thermoelectric effect can be estimated taking $X\rightarrow1$, where Andreev reflection is completely suppressed as one of the spin species vanishes. The finite but small thermoelectric effect at $X=1$ is thus only due to quasiparticle transfers.}. 
	
	\section{Odd-frequency pairing\label{sec:odd-w}}
	
	We now show how a finite thermoelectric current is only possible if even and odd-frequency Cooper pairs coexist with similar amplitude. 
	The Andreev reflection amplitude is related to the proximity-induced pairing, which can be described by the anomalous part of the Green's function associated to \cref{eq:hamil}. 
	Following Refs.~\cite{McMillan_1968,Furusaki_1991,Kashiwaya_1996,Kashiwaya_RPP,Asano_2001,Herrera_2010,Crepin_2015,Burset_2015,Kashuba_2017,Cayao_2017,Cayao_2018,Lu_2018,Cayao_2020} (see all the details in the appendix), we can write the anomalous Green's function as 
	\begin{equation}
	\hat{G}^r_{he}(x,x';E,\theta_S)=(f_0\si-f_3\sz)(i\sy),
	\end{equation} where we have identified the spin-symmetric singlet and triplet components $f_0$ and $f_3$, respectively, as
	\begin{equation}\label{eq:anom-GF-ST}
	f_{0,3}^r(0,0;E,\theta_S) = \frac{a_{S,T}}{i\hbar} = \frac{a_{1\up}(E,\theta_S)}{i2\hbar v_\up} \mp \frac{a_{1\dw}(E,\theta_S)}{i2\hbar v_\dw} .
	\end{equation}
	Here, $a_{S,T}$ are the singlet and triplet Andreev reflection amplitudes ($a_{1\sigma}$ corresponds to an injected electron in spin channel $\sigma$) and we have approximated the wave vectors as $k^\sigma_{e,h} \!\approx\! k_\sigma \pm E/(\hbar v_\sigma)$, with $k_\sigma\!=\!k_\text{F}\sqrt{\cos^2\theta_S+\rho X}$ and $v_\sigma\!=\!\hbar k_\sigma/m$. 
	It simplifies the symmetry analysis to use as reference the angle in the superconductor, $\theta_S$, which is the same for both spin channels, and only consider the anomalous Green's function evaluated at the interface, where $x=x'=0$. 
	
	Next, we define the even and odd parity parts of the anomalous functions as $f_{\mu,\pm}^{r}(E)\!=\![f_\mu^{r}(x,x',E,\theta_S)\pm f_\mu^{r}(x',x,E,-\theta_S)]/2$, with $\mu=0,3$ and the sign change in the angle due to the exchange $y\leftrightarrow y'$. 
	Then, to study the frequency dependence of the anomalous function, we use the retarded and advanced functions for positive and negative frequencies, respectively~\cite{Burset_2015}, since only there these functions have physical meaning. 
	Consequently, the anomalous function
	\begin{equation}\label{eq:f-freq-sim}
	f(x,x';E,\theta_S)= \left\{ \begin{array}{cr} f^{r}(x,x';E,\theta_S), &  E>0 \\  f^{a}(x,x';E,\theta_S), &  E<0  \end{array} \right. 
	\end{equation}
	connects both energy branches. Once the anomalous function in \cref{eq:f-freq-sim} is fully symmetric with respect to spin and parity, it belongs to one of four possible symmetry classes. 
	We label the different classes by their spin state, S for singlet and T for triplet, and their symmetry, even (E) or odd (O), with respect to frequency and space coordinates. For example, even-frequency, singlet, even-parity is ESE and ETO refers to the even-frequency, triplet odd-parity case~\cite{Golubov_2007,Golubov_2007b,Golubov_2007c,Tanaka_JPSJ}. 
	
	It is now useful to define the short-hand notation $a^{\alpha}_{\beta}=(\alpha E, \beta \theta)$, with $\alpha,\beta=\pm$ and $a$ standing for either of $a_{S,T}$, and thus write the energy- and angle-symmetric parts of the amplitude $a$ as
	\begin{subequations}\label{eq:EOPM-short}
		\begin{align}
		a^{Ee,Eo} ={}& \frac{1}{4} \left( a^{+}_{+} + a^{-}_{+} \pm a^{+}_{-} \pm a^{-}_{-} \right) , \\
		a^{Oe,Oo} ={}& \frac{1}{4} \left( a^{+}_{+} - a^{-}_{+} \pm a^{+}_{-} \mp a^{-}_{-} \right) . 
		\end{align}
	\end{subequations}
	Here, the index $a^{E,O}$ ($a^{e,o}$) labels the even and odd parts of the amplitude with respect to the energy (angle). 
	From the particle-hole symmetry of the Hamiltonian in \cref{eq:hamil} and the micro-reversibility of the scattering matrix, we obtain the condition 
	$v_{\dw,\up} a_{1\up,\dw}(E,\theta) =-v_{\up,\dw} a^*_{1\dw,\up}(-E,-\theta)$ for $|\theta_S|\leq\theta_c\!=\!\sin^{-1}\sqrt{1-X}$, see the appendix for more details. Once applied to the singlet and triplet amplitudes, this condition reads $a_{S,T}(E,\theta)=\pm a_{S,T}^*(-E,-\theta)$, and combined with \cref{eq:EOPM-short} reveals that, in the absence of virtual Andreev processes ($|\theta_S|\leq\theta_c$), the amplitudes $a_S^{Ee}$, $a_S^{Oo}$, $a_T^{Eo}$ and $a_T^{Oe}$ are real, while the rest are purely imaginary. 
	Therefore, at the interface we find
	\begin{subequations}\label{eq:pair-S}
		\begin{align}
		f_\text{ESE} ={}& \left\{ \begin{array}{rclr}
		-\frac{i}{2\hbar}(a_S^{Ee}+i\bar{a}_S^{Oe}) , & E>0 \\
		-\frac{i}{2\hbar}(a_S^{Ee}+i\bar{a}_S^{Oe}) , & E<0 \end{array} \right. , 
		\\
		f_\text{OSO} ={}& \left\{ \begin{array}{rclr}
		-\frac{i}{2\hbar}(a_S^{Oo}+i\bar{a}_S^{Eo}) , & E>0 \\
		+\frac{i}{2\hbar}(a_S^{Oo}+i\bar{a}_S^{Eo}) , & E<0 \end{array} \right. , 
		\\
		f_\text{ETO} ={}& \left\{ \begin{array}{rclr}
		-\frac{i}{2\hbar}(a_T^{Eo}+i\bar{a}_T^{Oo}) , & E>0 \\
		-\frac{i}{2\hbar}(a_T^{Eo}+i\bar{a}_T^{Oo}) , & E<0 \end{array} \right. , 
		\\
		f_\text{OTE} ={}& \left\{ \begin{array}{rclr}
		-\frac{i}{2\hbar}(a_T^{Oe}+i\bar{a}_T^{Ee}) , & E>0 \\
		+\frac{i}{2\hbar}(a_T^{Oe}+i\bar{a}_T^{Ee}) , & E<0 \end{array} \right. ,
		\end{align}
	\end{subequations}
	where we substituted the purely imaginary amplitudes as $a=i\bar{a}$, with $\bar{a}$ being real. It is then straightforward to check that $f_\text{ESE}$ and $f_\text{OSO}$ in \cref{eq:pair-S} are, respectively, even and odd in both frequency and parity~\footnote  {In the one-dimensional case with $\theta=0$, \cref{eq:pair-S} reproduces the pairing symmetries obtained in Ref.~\onlinecite{Cayao_2017}. }. 
	
	We can now establish a connection between the thermoelectric current and the symmetry of the pairing state. Focusing on the Andreev contribution to \cref{eq:ch-current_sym}, which we have demonstrated to be equal to the total current, we find
	\begin{gather}
	I_\text{A} =\frac{1}{eR_\text{N}}\int_{0}^{\infty}\! dE\, \delta f \nonumber \\ \times
	\mean{ v_\up v_\dw \left[ |a_S(E)|^2-|a_S(-E)|^2 
		+ |a_T(E)|^2-|a_T(-E)|^2 
		\right. \nonumber \\ \left.
		+ 2X\mathrm{Re}\left\{ a_S(E)a_T^*(E) - a_S(-E)a_T^*(-E) \right\} \right] 
	}_{\mbf{k}_\parallel} , 
	\label{eq:current-ST}
	\end{gather}
	where we have used that $\lambda_{\sigma}=v_{\bar{\sigma}}/v_\sigma$. 
	It is clear from \cref{eq:current-ST} that a finite polarization ($X\!\neq\!0$) couples the spin singlet and triplet states, resulting in a new contribution to the current. 
	To identify the scattering amplitudes with the anomalous Green's function, we first note that the virtual Andreev processes do not contribute to the current and thus the angle average is limited to $|\theta_S|\!\leq\!\theta_c$. 
	Then, using \cref{eq:EOPM-short,eq:pair-S} we can write the scattering amplitudes in the absence of virtual processes as 
	\begin{subequations}\label{eq:f-amp-conection_0}
		\begin{align}
		a_S(E)={}& 2i\hbar f_\text{ESE} + 2i\hbar f_\text{OSO} , 
		\\
		a_S(-E)={}& 2i\hbar \left(f_\text{ESE}\right)^* - 2i\hbar \left(f_\text{OSO}\right)^* , 
		\\
		a_T(E)={}& 2i\hbar f_\text{OTE} + 2i\hbar f_\text{ETO} , 
		\\
		a_T(-E)={}& 2i\hbar  \left(f_\text{ETO}\right)^* -2i\hbar \left(f_\text{OTE}\right)^* .
		\end{align}	
	\end{subequations}
	Inserting Eqs. (\ref{eq:f-amp-conection_0}) into \cref{eq:current-ST}, the current becomes
	\begin{gather}
	I_\text{A} = 
	\frac{16\hbar^2}{eR_\text{N}} \int_{0}^{\infty}\! dE\, \delta f \notag \\ \times
	\mathrm{Re} 
	\left\{ \mean{ v_\up v_\dw \left[
		\left(f_\text{ESE}\right)^* f_\text{OSO} + \left(f_\text{ETO}\right)^* f_\text{OTE} \right]
		\right. \notag \\ \left. +
		v_\up v_\dw X \left[ \left(f_\text{ESE}\right)^* f_\text{OTE} + \left(f_\text{ETO}\right)^* f_\text{OSO} \right] }_{\mbf{k}_\parallel} \right\} . 
	\label{eq:current-fEO}
	\end{gather}
	In general, the angle-average of the product of functions of opposite parity in the same spin state is zero, i.e., $\mean{f_\text{ESE}^*f_\text{OSO}}_{\mbf{k}_\parallel}=\mean{f_\text{ETO}^*f_\text{OTE}}_{\mbf{k}_\parallel}=0$. 
	Consequently, there is no thermoelectric current in \gls{nis} junctions, with $X=0$, even though a strong odd-frequency pairing appears in the presence of \gls{abs}~\cite{Tanaka_JPSJ}. 
	By contrast, in \gls{fis} junctions for arbitrary pairings, see \cref{fig:2dpair_param}, the thermoelectric effect is then due to the term proportional to $X$, which mixes even- and odd-frequency terms with different spin states~\footnote
	{The current, however, is not simply proportional to $X$, since the anomalous amplitudes also depend on the exchange field.}. 
	Therefore, the Andreev thermoelectric effect is only possible if even and odd-frequency pairing coexist~\cite{SunYong_2018}. 

	\begin{figure*}
		\includegraphics[width=0.90\textwidth]{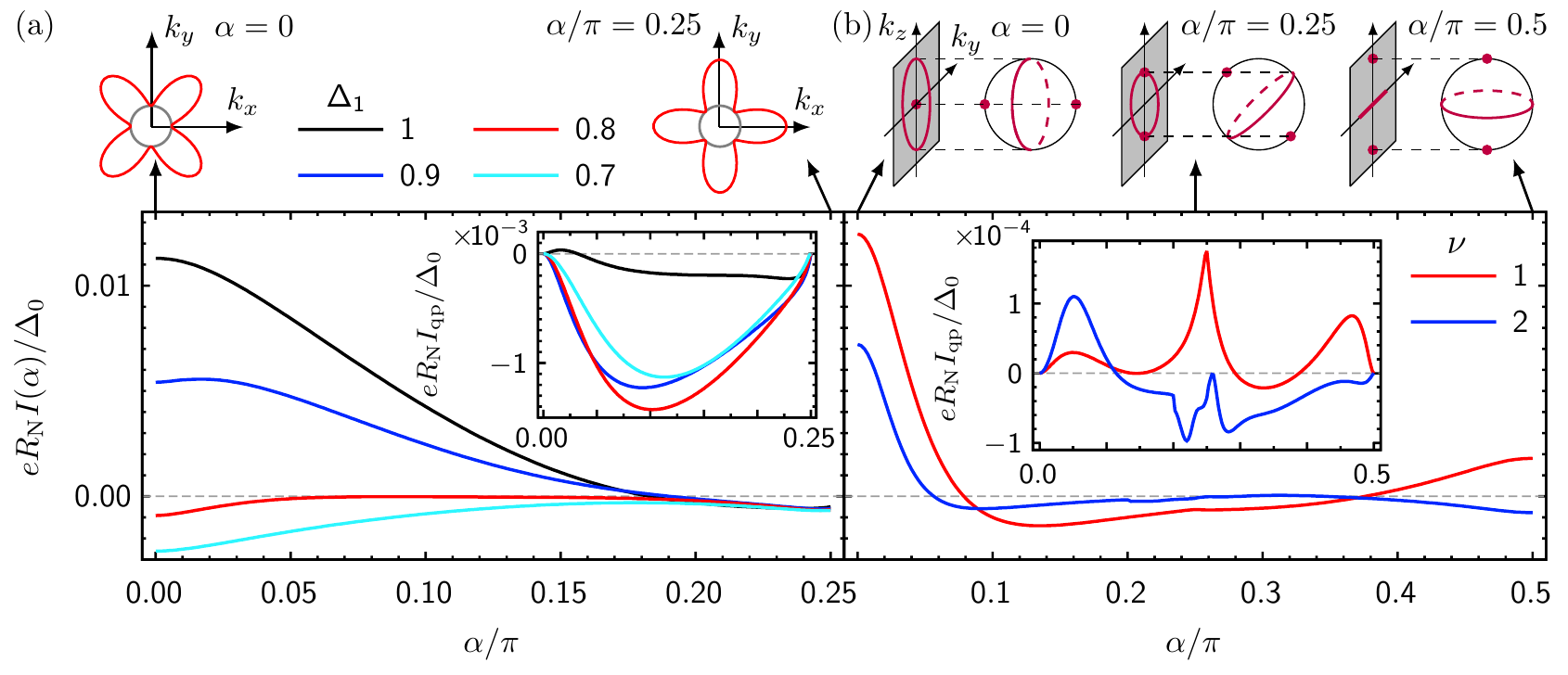}
		\caption{\label{fig:alpha-2d-3d} 
			Thermoelectric current as a function of the misalignment angle $\alpha$ in an \gls{fis} junction. 
			(a) Two-dimensional ($d_{xy}+is$)-wave pairing for different values of $\Delta_1$. 
			(b) 3D chiral pairing in \cref{eq:3d-pair} for $\nu=1,2$. 
			Top panels show the relative orientation between the pairing potential and the interface. 
			The insets show the quasiparticle contribution to the current. 
			In all cases we set $X=0.5$, $Z_0=3$, $Z_m\!=\!0$, $T=0.1T_\text{c}$, and $\delta T=0.2T_\text{c}$. 
		}
	\end{figure*}

	\section{Broken rotational symmetry\label{sec:qp-cont}}
	
	In all the previous results, the Andreev thermoelectric effect is due to subgap processes where only Cooper pairs are transmitted into the superconductor, see \cref{eq:current-andreev}, for any value of $X$ or $Z_m$. 
	However, the transport properties of unconventional anisotropic pairings depend strongly on the relative orientation between the pairing potential and the junction interface. 
	For example, for $d_{xy}$-wave pairing, the effective pairings for electron- and hole-like quasiparticles are $\Delta_\sigma^{\pm}=\pm\rho\Delta_0\sin\left(2\theta-2\alpha\right)$, respectively, where $\alpha$ is the misalignment angle~\cite{Tanaka_1995,Tanaka_1996}. 
	When $\alpha$ is different from one of the highly symmetric orientations, $\alpha\neq0,\pi/4$, the rotational symmetry for $d_{xy}$-wave is broken and $|\Delta_\sigma^+|\neq|\Delta_\sigma^-|$, resulting in a small contribution to the thermoelectric current from quasiparticles. 
	
	We show the effect of the misalignment angle in an \gls{fis} junction with $d_{xy}+is$-wave pairing in \cref{fig:alpha-2d-3d}(a). 
	When $\alpha\neq0,\pi/4$, the transport properties change. For normalized energies $E/\Delta_0$ below $\Delta_2=1-\Delta_1$, the system is fully gapped, the condition $\sigma= 2R_\text{N}\mean{P_\up A_\up+P_\dw A_\dw}_{\mbf{k}_\parallel}$ holds and the antisymmetric part of the conductance is only due to Andreev processes. 
	However, for $\Delta_2<|E/\Delta_0|<1$ quasiparticles can be transferred into the superconductor. 
	Consequently, the subgap condition $A_\sigma+B_\sigma=1$ no longer holds and the quasiparticle term $I_\text{qp}$ becomes finite in \cref{eq:ch-current}. 
	Even though the current now has two contributions, $I_\text{A}$ and $I_\text{qp}$, we are clearly in the situation where $I_\text{A}>I_\text{qp}$, see inset of \cref{fig:alpha-2d-3d}(a). 
	For the two-dimensional pairings considered here, the quasiparticle current generated by a broken rotational symmetry is not enough to change the sign of the total current around the highly symmetric orientations. 
	
	The results from the previous section can be extended to 3D unconventional superconductors without loss of generality. In higher dimensions, however, the energy dispersion of \gls{abs} can become more complicated~\cite{Murakawa_2009,Sasaki_2011}. 
	Recently, the exotic \gls{abs} of 3d chiral superconductors have attracted a lot of attention due to the simultaneous presence of line and point nodes~\cite{Asano_2003,Qi_2009,Chung_2009,Fu_2010,Hao_2011,Hsieh_2012,Yamakage_2012,Hashimoto_2015,Bo_2015,Hashimoto_2016}. 
	The \gls{abs} of 3d chiral superconductors are predicted to originate from the pair potential
	\begin{equation}\label{eq:3d-pair}
	\hat{\Delta}^\nu_{3\text{d}}(\mbf{k}) = \Delta_\text{T} \frac{k'_x\left(k_y+ik'_z\right)^\nu}{r_\nu k_\text{F}^{\nu+1}} \times \left\{ 
	\begin{array}{cl}
	i\sy  , & \nu \text{ odd} \\
	\sx , & \nu \text{ even} 
	\end{array}
	\right. ,
	\end{equation}
	with $k'_{x}=k_{x}\cos\alpha- k_{z}\sin\alpha$, $k'_{z}=k_{x}\sin\alpha+ k_{z}\cos\alpha$, $\alpha$ being the misalignment angle from the $k_x$ axis, $\nu$ a nonzero integer~\cite{Kobayashi_2015,Tamura_2017}, and the normalization factors $(r_{\nu=0},r_{\nu=1},r_{\nu=2})=(1,1/2,2/\sqrt{27})$. 
	For $\nu=0$, the potential features a line node, while it has two point nodes and a line node for $\nu=1,2$. The pairing is in a spin-triplet (singlet) state for $\nu$ even (odd). 
	
	While the case with $\nu=0$ generalizes the two-dimensional chiral $p$-wave triplet pairing of the previous section, the cases with finite $\nu$ are relevant for heavy fermion superconductors, with $\nu=1$ and $\nu=2$ corresponding to the candidate pairing symmetries of URu$_2$Si$_2$ \cite{Matsuda_2007,Matsuda_2014,Kapitulnik_2015} and UPt$_3$ \cite{Joynt_2002,Ozaki_2012,Machida_2013,Schemm_2014,Goswami_2015}, respectively. The topological origin of the flat-band \gls{abs} and its fragility against surface misalignment have been clarified in previous works~\cite{Kobayashi_2015}, together with the \gls{abs} energy dispersion~\cite{Tamura_2017}. 
	It was found that a unique characteristic of the $\nu\!=\!1,2$ pairings is that an external exchange field can turn a zero-bias conductance peak into a dip, due to the complicated dispersion of the \gls{abs}~\cite{Tamura_2017}. 
	Consequently, an equivalent sign change in the thermoelectric current appears here. 
	
	The thermoelectric current for 3d chiral superconductors behaves like the two-dimensional cases for highly symmetric orientations ($\alpha=0,\pi/2$). For different surface misalignments, however, the quasiparticle contribution from broken rotational symmetry can affect the sign of the current, see \cref{fig:alpha-2d-3d}(b). 
	In the presence of a zero-energy \gls{abs}, however, the quasiparticle contribution is smaller than the Andreev one and the current remains positive. 
	The sign change of the thermoelectric current thus clearly indicates the presence of \gls{abs}, even when they have a complicated energy dispersion. Moreover, some orientations, like $\alpha=\pi/2$, have a very different zero-energy behavior depending on the value of $\nu$. For $\nu=1$ and $\nu=2$, the pairing can be regarded as a chiral $p$-wave or chiral $d$-wave, which feature the opposite signs of the current. Consequently, the sign of the thermoelectric current is, again, a good indicator of the presence of \gls{abs}, even if the surface states display a complicated dispersion relation. 
	
	\section{Summary\label{sec:conc}}
	
	Following a microscopic scattering approach, we have shown that the thermoelectric effect in ferromagnet-superconductor junctions can be entirely dominated by subgap Andreev processes. Consequently, the electric current from a temperature bias changes sign in the presence of \gls{abs} from unconventional superconductivity and is thus a sensitive tool to identify these surface states. 
	
	When the rotational symmetry of the pairing potential is maintained at the junction interface (that is, $|\Delta_\sigma^+|=|\Delta_\sigma^-|$ for each spin channel $\sigma$) the transport asymmetry that gives rise to the thermoelectric effect occurs only at subgap energies. The resulting current is dominated by Andreev processes, $I(\delta T)=2I_\text{A}(\delta T)$, with no quasiparticle transfer into the superconductor. A finite polarization induced by a Zeeman exchange field $X$ in an \gls{fis} junction is a necessary and sufficient condition for this Andreev thermoelectric effect. An additional intermediate ferromagnetic insulator, \gls{ffis} junction, enhances the magnitude of the current if the majority spin species is favored (i.e., $Z_m>0$ for $X>0$). We have shown how the exchange field allows for the coupling of even- and odd-frequency Cooper pairs leaking out of the superconductor, which provide the main contribution to the Andreev thermoelectric current. 
	Thermoelectric measurements in \gls{fis} setups can thus complement recent experimental signatures of odd-frequency pairing~\cite{Simon_2020,*Simon_2020b}. 

	For \gls{ffis} junctions, we restricted our analysis to the case of collinear exchange fields in the ferromagnet and ferromagnetic insulator. For non-collinear fields, a long-range, odd-frequency triplet state emerges~\cite{Bergeret_2001,*Bergeret_RMP,Birge_2010,Robinson_2010,Robinson_2010a}, which could enhance the thermoelectric current according to \cref{eq:current-fEO}~\cite{Dutta_2017,Dutta_2020}. 
	In particular, Ref.~[\onlinecite{Dutta_2020}] shows that a large figure of merit is possible for \gls{ffis} junctions with non-collinear exchange fields and conventional superconductors. 
	
	When the pairing rotational symmetry is broken ($|\Delta_\sigma^+|\!\neq\!|\Delta_\sigma^-|$), quasiparticle transfers into the superconductor are allowed and contribute to the thermoelectric current. Such contributions could change the sign of the current and mask the presence of \gls{abs}. 
	However, we have analyzed the thermoelectric effect for two- and three-dimensional pairings with a strong angular dependence and when low-energy \gls{abs} are present the current is positive, Andreev-dominated and with a strong magnitude around highly-symmetric orientations. 
	As such, the sign of the thermoelectric current could help identify the pairing state even for cases where the outcome of tunneling spectroscopy can be ambiguous. 
	
	In conclusion, exploring the thermoelectric effect in ferromagnet-superconductor hybrid junctions involving unconventional superconductors offers a sensitive method to identify emerging surface Andreev states. Our results can be easily extended to heterostructures where topological superconductivity is artificially engineered by combining conventional superconductors with topological materials~\cite{Jack_2019}.

	\acknowledgments
	We thank A.~Black-Schaffer and B.~Sothmann for insightful discussions. 
	We acknowledge support from the Academy of Finland (projects No.~308515 and 312299), Topological Material Science (Grants No.~JP15H05851, No.~JP15H05853, and No.~JP15K21717), Scientific Research A (KAKENHI Grant No.~JP20H00131), JSPS Core-to-Core program “Oxide Superspin international network", Japan-RFBR Bilateral Joint Research Projects/Seminars No. 19-52-50026, and Grant-in-Aid for Scientific Research B (KAKENHI Grant No. JP18H01176 and No. JP20H01857) from the Ministry of Education, Culture, Sports, Science, and Technology, Japan (MEXT), and the Spanish CAM "Talento Program" No. 2019-T1/IND-14088. 
	
	\appendix
	
	\begin{widetext}
		
		\section{Retarded Green's function}
		
		Following Refs.~\cite{McMillan_1968,Furusaki_1991,Kashiwaya_1996,Kashiwaya_RPP,Asano_2001,Herrera_2010,Crepin_2015,Burset_2015,Kashuba_2017,Cayao_2017,Cayao_2018,Lu_2018,Cayao_2020}, the retarded Green's function associated with the Hamiltonian in \cref{eq:hamil} is constructed from a linear combination of scattering states obtained from \cref{eq:gen-sol} with the appropriate boundary conditions. We now describe the main steps of this method for a generic pairing potential. 
		
		First, since the spin channels in \cref{eq:hamil} are decoupled, we only need to work with the reduced Hamiltonian
		\begin{equation}
		\hat{H}_\sigma(x,\mathbf{k})=\left(\begin{array}{cc}
		h_\sigma(x) - \mu & \Delta_\sigma(\mathbf{k}_\parallel/k_\text{F}) \e^{i\phi} \\ \Delta^*_\sigma(\mathbf{k}_\parallel/k_\text{F}) \e^{-i\phi} & \mu - h_\sigma(x)
		\end{array}\right) ,
		\label{eq:hamil-spin}
		\end{equation}
		where
		\begin{equation}\label{eq:hamil-sp-scalar}
		h_\sigma(x)= -\frac{\hbar^2}{2m} \frac{d^2}{dx^2} -\rho U(x) + \left(V_0 -\rho U_\text{B}\right) \delta(x) .
		\end{equation}
		The retarded Green's function thus satisfies the equation of motion 
		\begin{equation}
		\delta(x-x') = \left[\hat{E}_+ - \hat{H}_\sigma(x,\mathbf{k})\right]\hat{G}^r_\sigma(x,x';E,\mbf{k}_\parallel) 
		= \hat{G}^r_\sigma(x,x';E,\mbf{k}_\parallel) \left[\hat{E}_+ - \hat{H}^T_\sigma(x,\mathbf{k})\right] ,
		\end{equation}
		with $\hat{E}_+=(E+i0^+)\si$, which can be integrated to obtain the following boundary conditions:
		\begin{subequations}\label{eq:GF-bbcc}
			\begin{align}
			0={}& \hat{G}^r_\sigma(x,x+0^+) - \hat{G}^r_\sigma(x,x-0^+) , \\
			\frac{2m}{\hbar^2} \tz ={}& \left. \partial_x \hat{G}^r_\sigma(x,x') \right|_{x=x'+0^+} - \left. \partial_x \hat{G}^r_\sigma(x,x') \right|_{x=x'-0^+} .
			\end{align}
		\end{subequations}
		
		A general solution of the \gls{bdg} equations for $\hat{H}_\sigma(x,\mathbf{k})$ takes the form
		\begin{equation}
		\Phi_\sigma(\mbf{r}) = \e^{i ( k_y y + k_z z) } \sum\limits_{\alpha=\pm} \left[ 
		m^\alpha_\sigma \psi_{\alpha k^\sigma_{1}} \e^{\alpha i k^\sigma_{1} x}
		+
		n^\alpha_\sigma \psi_{\alpha k^\sigma_{2}} \e^{ \alpha i k^\sigma_{2} x}
		\right] , \notag
		\end{equation}
		where $\psi_{\pm k^\sigma_{1}} = [1, (\eta^{\pm}_\sigma)^*\Gamma^\pm_\sigma]^T$ ($\psi_{\pm k^\sigma_{2}} = [\eta^{\pm}_\sigma\Gamma^\pm_\sigma, 1]^T$) is the solution for an electron-like (hole-like) quasiparticle with amplitude $m^\alpha_\sigma$ ($n^\alpha_\sigma$), cf. \cref{eq:gen-sol}. 
		For the transposed Hamiltonian, we find 
		\begin{equation}
		\tilde{\Phi}_\sigma(\mbf{r}) = \e^{-i ( k_y y + k_z z) } \sum\limits_{\alpha=\pm} \left[ 
		\tilde{m}^\alpha_\sigma \tilde{\psi}_{\alpha k^\sigma_{1}} \e^{ \alpha i k^\sigma_{1} x}
		+
		\tilde{n}^\alpha_\sigma \tilde{\psi}_{\alpha k^\sigma_{2}} \e^{ \alpha i k^\sigma_{2} x}
		\right] , \notag
		\end{equation}
		with $\tilde{\psi}_{\pm k^\sigma_{1}} = [1, \eta^{\mp}_\sigma\Gamma^\mp_\sigma]^T$ and $\tilde{\psi}_{\pm k^\sigma_{2}} =  [(\eta^{\mp}_\sigma)^*\Gamma^\mp_\sigma, 1]^T$. 
		Notice that both the extra conjugation and the exchange $\theta_+\leftrightarrow\theta_-$ are a result of the transposition (we take the superconducting phase to be zero for the sake of simplicity). 
		The four possible scattering states for an \gls{fis} junction are injection of electron (1) or hole (2) from the ferromagnet and injection of an electron- (3) or hole-like (4) quasiparticle from the superconductor. Specifically, we have
		\begin{subequations}
			\begin{align}
			\phi_{\sigma,1}(x)={}& \left\{ \begin{array}{cr}
			\psi_{+k^\sigma_{e}}\e^{i k^\sigma_{e} x} 
			+ a^\sigma_{1} \psi_{+k^\sigma_{h}}\e^{ik^\sigma_{h} x} 
			+ b^\sigma_{1} \psi_{-k^\sigma_{e}}\e^{-ik^\sigma_{e} x}
			, & x<0 \\ 
			c^\sigma_{1} \psi_{+k^\sigma_{1}}\e^{ik^\sigma_{1} x} 
			+ d^\sigma_{1} \psi_{-k^\sigma_{2}}\e^{-ik^\sigma_{2} x}, & x>0 
			\end{array} \right. ,
			\\ 
			\phi_{\sigma,2}(x)={}& \left\{ \begin{array}{cr}
			\psi_{-k^\sigma_{h}}\e^{-ik^\sigma_{h} x}
			+ b^\sigma_{2} \psi_{+k^\sigma_{h}}\e^{ik^\sigma_{h} x} 
			+ a^\sigma_{2} \psi_{-k^\sigma_{e}}\e^{-ik^\sigma_{e} x}
			, & x<0 \\ 
			d^\sigma_{2} \psi_{+k^\sigma_{1}}\e^{ik^\sigma_{1} x} 
			+ c^\sigma_{2} \psi_{-k^\sigma_{2}}\e^{-ik^\sigma_{2} x}, & x>0 
			\end{array} \right. ,
			\\ 
			\phi_{\sigma,3}(x)={}& \left\{ \begin{array}{cr}
			d^\sigma_{3} \psi_{+k^\sigma_{h}}\e^{ik^\sigma_{h} x}
			+ c^\sigma_{3} \psi_{-k^\sigma_{e}}\e^{-ik^\sigma_{e} x}
			, & x<0 \\ 
			\psi_{-k^\sigma_{1}}\e^{-ik^\sigma_{1} x}
			+ b^\sigma_{3} \psi_{+k^\sigma_{1}}\e^{ik^\sigma_{1} x}
			+ a^\sigma_{3} \psi_{-k^\sigma_{2}}\e^{-ik^\sigma_{2} x}, & x>0 
			\end{array} \right. ,
			\\ 
			\phi_{\sigma,4}(x)={}& \left\{ \begin{array}{cr}
			c^\sigma_{4} \psi_{+k^\sigma_{h}}\e^{ik^\sigma_{h} x}
			+ d^\sigma_{4} \psi_{-k^\sigma_{e}}\e^{-ik^\sigma_{e} x}
			, & x<0 \\ 
			\psi_{k^\sigma_{2}}\e^{ik^\sigma_{1} x} 
			+ a^\sigma_{4} \psi_{+k^\sigma_{1}}\e^{ik^\sigma_{1} x}
			+ b^\sigma_{4} \psi_{-k^\sigma_{2}}\e^{-ik^\sigma_{2} x}, & x>0 
			\end{array} \right. ,
			\end{align}
		\end{subequations}
		where the wave vectors in the normal regions are $k^\sigma_{e,h}$ and thus $\psi_{\pm k^\sigma_{e}}\!=\!(1,0)^T$ and $\psi_{\pm k^\sigma_{h}}\!=\!(0,1)^T$, since $\eta^{\pm}_\sigma\!=\!0$. 
		Finally, the retarded Green's function is a linear combination of asymptotic solutions of $\hat{H}_\sigma(x,\mathbf{k})$ and $\hat{H}^T_\sigma(x,\mathbf{k})$, namely,
		\begin{equation}\label{eq:GF-def}
		\hat{G}^r_\sigma(x,x') = 
		\left\{ \begin{array}{lr} 
		\phi_{\sigma,3}(x) \left[\alpha^\sigma_1 \tilde{\phi}_{\sigma,1}^{T}(x') + \alpha^\sigma_3 \tilde{\phi}_{\sigma,2}^{T}(x') \right]
		+\phi_{\sigma,4}(x) \left[\alpha^\sigma_2 \tilde{\phi}_{\sigma,1}^{T}(x') + \alpha^\sigma_4 \tilde{\phi}_{\sigma,2}^{T}(x') \right] 
		, & x<x' \\
		\phi_{\sigma,1}(x) \left[\beta^\sigma_1 \tilde{\phi}_{\sigma,3}^{T}(x') + \beta^\sigma_3 \tilde{\phi}_{\sigma,4}^{T}(x') \right] 
		+\phi_{\sigma,2}(x) \left[\beta^\sigma_2 \tilde{\phi}_{\sigma,3}^{T}(x') + \beta^\sigma_4 \tilde{\phi}_{\sigma,4}^{T}(x') \right] 
		, & x>x' 
		\end{array}
		\right. . 
		\end{equation}
		
		The coefficients of the linear combination are obtained by imposing the boundary conditions in \cref{eq:GF-bbcc}. Focusing on the region with $x,x'<0$, we find from the Nambu-diagonal elements the system of equations
		\begin{subequations}
			\begin{align}
			d^\sigma_3 \alpha^\sigma_1 + c^\sigma_4 \alpha^\sigma_3 = 
			\tilde{d}^\sigma_3 \beta_1 + \tilde{c}^\sigma_4 \beta^\sigma_3 =
			c^\sigma_3 \alpha^\sigma_2 + d^\sigma_4 \alpha^\sigma_4 = 
			\tilde{c}^\sigma_3 \beta_2 + \tilde{d}^\sigma_4 \beta^\sigma_4 ={}& 0 , \\
			c^\sigma_3 \alpha^\sigma_1 + d^\sigma_4 \alpha^\sigma_3 - \tilde{c}^\sigma_3 \beta_1 - \tilde{d}^\sigma_4 \beta^\sigma_3 =
			d^\sigma_3 \alpha^\sigma_2 + c^\sigma_4 \alpha^\sigma_4 - \tilde{d}^\sigma_3 \beta_2 - \tilde{c}^\sigma_4 \beta^\sigma_4 ={}& 0 , \\ 
			c^\sigma_3 \alpha^\sigma_1 + d^\sigma_4 \alpha^\sigma_3 + \tilde{c}^\sigma_3 \beta_1 + \tilde{d}^\sigma_4 \beta^\sigma_3 ={}& \frac{2m}{i\hbar^2k^\sigma_e} , \\ 
			d^\sigma_3 \alpha^\sigma_2 + c^\sigma_4 \alpha^\sigma_4 + \tilde{d}^\sigma_3 \beta_2 + \tilde{c}^\sigma_4 \beta^\sigma_4 ={}& \frac{2m}{i\hbar^2k^\sigma_h} , 
			\end{align}
		\end{subequations}
		resulting in
		\begin{subequations}\label{eq:lin-coefs}
			\begin{align}
			\alpha^\sigma_1 ={}& -\frac{c^\sigma_4}{d^\sigma_3} \alpha^\sigma_3 = \frac{-i}{\hbar v^\sigma_{e} } \frac{c^\sigma_4}{c^\sigma_3c^\sigma_4-d^\sigma_3d^\sigma_4} , 
			\quad 
			\alpha^\sigma_2 = -\frac{d^\sigma_4}{c^\sigma_3} \alpha^\sigma_4 = \frac{i}{\hbar v^\sigma_{h} } \frac{d^\sigma_4}{c^\sigma_3c^\sigma_4-d^\sigma_3d^\sigma_4} , 
			\\ 
			\beta^\sigma_1 ={}& -\frac{\tilde{c}^\sigma_4}{\tilde{d}^\sigma_3} \beta^\sigma_3 = \frac{-i}{\hbar v^\sigma_{e} } \frac{\tilde{c}^\sigma_4}{\tilde{c}^\sigma_3\tilde{c}^\sigma_4-\tilde{d}^\sigma_3\tilde{d}^\sigma_4} , 
			\quad
			\beta^\sigma_2 = -\frac{\tilde{d}^\sigma_4}{\tilde{c}^\sigma_3} \beta^\sigma_4 = \frac{i}{\hbar v^\sigma_{h} } \frac{\tilde{d}^\sigma_4}{\tilde{d}^\sigma_3\tilde{d}^\sigma_4-\tilde{d}^\sigma_3\tilde{d}^\sigma_4} , 
			\end{align}
		\end{subequations}
		with $v^\sigma_{e,h}=\hbar k^\sigma_{e,h}/m$ being the velocity for spin $\sigma$ electrons and holes, respectively. 
		On the other hand, the Nambu off-diagonal sector provides the detailed balance equations
		\begin{equation}\label{eq:det-bal}
		k^\sigma_h \tilde{a}^\sigma_1 = k^\sigma_e a^\sigma_2 , \quad 
		k^\sigma_h a^\sigma_1 = k^\sigma_e \tilde{a}^\sigma_2 ,
		\end{equation}
		which can be demonstrated using the Wronskian of the scattering processes 1 and 2. For Nambu spinors $\psi_j(x)\!=\![u_j(x),v_j(x)]^T$, the Wronskian is defined as
		\begin{equation}
		W\left[\psi_i(x),\psi_j(x)\right]\equiv u_i u'_j - u_j u'_i - \left( v_i v'_j - v_j v'_i \right) ,
		\end{equation}
		with $u'\!=\!du(x)/dx$. For the scattering processes 1 and 2, we obtain
		\begin{equation}\label{eq:det-bal1}
		W [\phi_{\sigma,1}(x),\tilde{\phi}_{\sigma,2} ] = \left\{ \begin{array}{cr}
		2i\left( k^\sigma_h a^\sigma_1 - k^\sigma_e \tilde{a}^\sigma_2 \right) , & x<0 \\ 0, & x>0
		\end{array} \right. .
		\end{equation}
		The Wronskian is independent of the coordinate $x$ and, thus, equating the two branches of the Wronskian, we obtain one of the detailed balance equations in \cref{eq:det-bal}. The other one comes from $W[\tilde{\phi}_{\sigma,1}(x),\phi_{\sigma,2}]$. 
		
		By plugging the linear coefficients in \cref{eq:lin-coefs} into \cref{eq:GF-def}, and using the detailed balance relations in \cref{eq:det-bal}, we obtain the retarded Green's function per spin channel in the normal region as
		\begin{equation}
		\hat{G}^r_\sigma(x,x')= \frac{1}{i\hbar} \begin{pmatrix}
		\frac{1}{v_e^\sigma} \left[ \e^{i k^\sigma_e |x-x'|} + b^\sigma_1 \e^{i k^\sigma_e (x+x')} \right] & 
		\frac{a^\sigma_2}{v_h^\sigma} \e^{-i( k^\sigma_e x - k^\sigma_h x') } \\
		\frac{a^\sigma_1}{v_e^\sigma} \e^{ i( k^\sigma_h x - k^\sigma_e x') } &
		\frac{1}{v_h^\sigma} \left[ \e^{-i k^\sigma_h |x-x'|} + b^\sigma_2 \e^{-i k^\sigma_h (x+x')} \right]
		\end{pmatrix} .
		\end{equation}
		
		\section{Symmetry analysis of the anomalous Green's function}
		
		Using as reference the angle in the superconductor, $\theta_S$, which does not change with the spin channel, we approximate the wave vectors in the normal region as follows
		\begin{equation}
		k^\sigma_{e(h)} = k_F\sqrt{1 + \rho X +(-)\frac{E}{\mu} - \frac{k^2_y}{k^2_F} } 
		= k_F\sqrt{\cos^2\theta_S + \rho X +(-)\frac{E}{\mu} } 
		\approx k_\sigma +(-)\frac{E}{\hbar v_\sigma} ,
		\end{equation}
		with $k_\sigma\!=\!k_F\sqrt{\cos^2\theta_S+\rho X}$ and $v_\sigma\!=\!\hbar k_\sigma/m$. 
		The spatial dependence of the anomalous part is thus approximated as 
		\begin{subequations}
			\begin{align}
			\frac{1}{v^\up_e} \e^{i( k^\up_h x - k^\up_e x') } ={}& 
			\frac{1}{v_\up} \e^{i( k_\dw x - k_\up x') } \e^{-i\frac{E}{\hbar}( \frac{x}{v_\dw} + \frac{x'}{v_\up} ) }
			, \quad &
			\frac{1}{v^\up_h} \e^{-i( k^\up_e x - k^\up_h x') } ={}& 
			\frac{1}{v_\dw} \e^{-i( k_\up x - k_\dw x') } \e^{-i\frac{E}{\hbar}( \frac{x}{v_\up} + \frac{x'}{v_\dw} ) } ,
			\\
			\frac{1}{v^\dw_e} \e^{i( k^\dw_h x - k^\dw_e x') } ={}&
			\frac{1}{v_\dw} \e^{i( k_\up x - k_\dw x') } \e^{-i\frac{E}{\hbar}( \frac{x}{v_\up} + \frac{x'}{v_\dw} ) } 
			, \quad &
			\frac{1}{v^\dw_h} \e^{-i( k^\dw_e x - k^\dw_h x') } ={}&
			\frac{1}{v_\up} \e^{-i( k_\dw x - k_\up x') } \e^{-i\frac{E}{\hbar}( \frac{x}{v_\dw} + \frac{x'}{v_\up} ) } .
			\end{align}
		\end{subequations}
		
		The anomalous part of the spinful Green's function is then given by the functions
		\begin{align}
		\hat{G}^r_{he}(x,x')={}& \frac{1}{i\hbar} \begin{pmatrix}
		0 & \frac{a^\dw_1}{v_\dw} \e^{i( k_\up x - k_\dw x') } \e^{-i\frac{E}{\hbar}( \frac{x}{v_\up} + \frac{x'}{v_\dw} ) } \\ 
		\frac{a^\up_1}{v_\up} \e^{i( k_\dw x - k_\up x') } \e^{-i\frac{E}{\hbar}( \frac{x}{v_\dw} + \frac{x'}{v_\up} ) } & 0 
		\end{pmatrix} ,
		\label{eq:anom-GF_ret}
		\\
		\hat{G}^r_{eh}(x,x')={}& \frac{1}{i\hbar} \begin{pmatrix}
		0 & \frac{a^\up_2}{v_\dw} \e^{-i( k_\up x - k_\dw x') } \e^{-i\frac{E}{\hbar}( \frac{x}{v_\up} + \frac{x'}{v_\dw} ) } \\ 
		\frac{a^\dw_2}{v_\up} \e^{-i( k_\dw x - k_\up x') } \e^{-i\frac{E}{\hbar}( \frac{x}{v_\dw} + \frac{x'}{v_\up} ) } & 0 
		\end{pmatrix} .
		\end{align}
		
		Next, we compute the advanced anomalous Green's function $\hat{G}^a_{he}$ from $\hat{G}^r_{eh}$ using the property $\check{G}^a(\mbf{r},\mbf{r}')\!=\![\check{G}^r(\mbf{r}',\mbf{r})]^\dagger$, resulting in
		\begin{equation}\label{eq:anom-GF_adv}
		\hat{G}^a_{he}(x,x')= -\frac{1}{i\hbar} \begin{pmatrix}
		0 & \frac{[a^\dw_2(E,-\theta_S)]^*}{v_\up} \e^{-i( k_\up x - k_\dw x') } \e^{i\frac{E}{\hbar}( \frac{x}{v_\up} + \frac{x'}{v_\dw} ) } \\ 
		\frac{[a^\up_2(E,-\theta_S)]^*}{v_\dw} \e^{-i( k_\dw x - k_\up x') } \e^{i\frac{E}{\hbar}( \frac{x}{v_\dw} + \frac{x'}{v_\up} ) } & 0 
		\end{pmatrix} .
		\end{equation}
		At the interface, $x\!=\!0$, the anomalous Green's functions reduce to
		\begin{equation}
		\hat{G}^r_{he}(0,0)= \frac{1}{i\hbar} \begin{pmatrix}
		0 & \frac{a^\dw_1(E,\theta_S)}{v_\dw}\\ 
		\frac{a^\up_1(E,\theta_S)}{v_\up}  & 0 
		\end{pmatrix} , \quad 
		\hat{G}^a_{he}(0,0)= -\frac{1}{i\hbar} \begin{pmatrix}
		0 & \frac{[a^\dw_2(E,-\theta_S)]^*}{v_\up} \\ 
		\frac{[a^\up_2(E,-\theta_S)]^*}{v_\dw}  & 0 
		\end{pmatrix} .
		\end{equation}
		From the detailed balance relations within a spin sector, see \cref{eq:det-bal1}, we find that, if $\alpha\!=\!0$, 
		\begin{equation}\label{eq:det-bal2}
		\left. \begin{array}{c}
		v_\dw a^{\up}_1(E,\theta_S)= \mp v_\up a^\up_2(E,-\theta_S) \\
		v_\up a^\dw_1(E,\theta_S)= \mp v_\dw a^\dw_2(E,-\theta_S)
		\end{array}\right\} \equiv \,
		v_{\bar{\sigma}} a^{\sigma}_1(E,\theta_S)= \mp v_{\sigma} a^\sigma_2(E,-\theta_S) ,
		\end{equation}
		with the minus sign for singlet and plus for triplet. Particle-hole symmetry, for $|\theta_S|\leq\theta_c\!=\!\sin^{-1}\sqrt{1-X}$, imposes that 
		\begin{equation}\label{eq:PHS}
		a^{\up,\dw}_1(E,\theta_S)=\pm [a^{\dw,\up}_2(-E,\theta_S)]^* ,
		\end{equation}
		which changes the spin sector. Applying this property to \cref{eq:anom-GF_adv} relates the retarded and advanced anomalous functions as $\hat{G}^a_{he}(x,x';E,\theta_S)\!=\!\mp [\hat{G}^r_{he}(x',x;-E,-\theta_S)]^T$. For $|\theta_S|\leq\theta_c$, we can combine both properties to obtain 
		\begin{equation}\label{eq:PHS-det-bal}
		v_{\dw,\up} a^{\up,\dw}_1(E,\theta_S)= - v_{\up,\dw} [a^{\dw,\up}_1(-E,-\theta_S)]^* ,\quad
		v_{\up,\dw} a^{\up,\dw}_2(E,\theta_S)= - v_{\dw,\up} [a^{\dw,\up}_2(-E,-\theta_S)]^* ,
		\end{equation}
		which connects the same scattering process for different spin sectors. Applying these relations to the anomalous Green's function at the interface, we obtain
		\begin{equation}
		\hat{G}^a_{he}[E,\theta_S]= 
		\frac{\pm1}{i\hbar} \begin{pmatrix}
		0 & \frac{(a^\dw_1)^*}{v_\dw} \\ 
		\frac{(a^\up_1)^*}{v_\up}  & 0 
		\end{pmatrix}[E,-\theta_S] =
		\frac{\mp1}{i\hbar} \begin{pmatrix}
		0 & \frac{a^\up_1}{v_\up}  \\ 
		\frac{a^\dw_1}{v_\dw}  & 0 
		\end{pmatrix}[-E,\theta_S] = 
		\frac{1}{i\hbar} \begin{pmatrix}
		0 & \frac{a^\up_2}{v_\dw} \\ 
		\frac{a^\dw_2}{v_\up}  & 0 
		\end{pmatrix}[-E,-\theta_S] .
		\end{equation}
		
		From \cref{eq:anom-GF_ret} and \cref{eq:anom-GF_adv}, we define the singlet and triplet components of the anomalous Green's functions as
		\begin{subequations}\label{eq:anom-GF-ST_FIS}
			\begin{align}
			f^r_{0,3}(x,x',E,\theta_S)={}& \frac{-i}{2\hbar} \left[
			\frac{a^\up_1(E,\theta_S)}{v_\up} \e^{i( k_\dw x - k_\up x') } \e^{-i\frac{E}{\hbar}( \frac{x}{v_\dw} + \frac{x'}{v_\up} ) } \mp 
			\frac{a^\dw_1(E,\theta_S)}{v_\dw} \e^{i( k_\up x - k_\dw x') } \e^{-i\frac{E}{\hbar}( \frac{x}{v_\up} + \frac{x'}{v_\dw} ) }
			\right] ,
			\\
			f^a_{0,3}(x,x',E,\theta_S)={}& \mp \frac{i}{2\hbar} \left[
			\frac{a^\up_1(-E,-\theta_S)}{v_\up}  \e^{-i( k_\up x - k_\dw x') } \e^{i\frac{E}{\hbar}( \frac{x}{v_\up} + \frac{x'}{v_\dw} ) } \mp 
			\frac{a^\dw_1(-E,-\theta_S)}{v_\dw}\e^{-i( k_\dw x - k_\up x') } \e^{i\frac{E}{\hbar}( \frac{x}{v_\dw} + \frac{x'}{v_\up} ) } 
			\right] .
			\end{align}
		\end{subequations}
		The retarded and advanced functions are related by the transformation $f_{0,3}^a(x,x';E,\theta_S)\!=\!\pm f_{0,3}^r(x',x;-E,-\theta_S)$, cf. \cref{eq:anom-GF-ST_FIS}, which connects the positive and negative energy branches, where these functions have physical meaning. 
		To investigate the frequency dependence of the anomalous function, we define a function equal to the retarded (advanced) anomalous function for positive (negative) frequencies~\cite{Burset_2015,Cayao_2017}. Assuming $\epsilon\!>\!0$, we write the energy as $E=\sgn(E)\epsilon$ to find
		\begin{equation}\label{eq:f-sym}
		f_{0,3}(x,x';E,\theta_S)= \left\{ \begin{array}{cr} f_{0,3}^r(x,x';\epsilon,\theta_S), &  E>0 \\  -f_{0,3}^a(x,x';-\epsilon,\theta_S), &  E<0  \end{array} \right. ,
		\end{equation}
		where the sign for negative energies takes into account the different spin-symmetrization of the advanced Green's function. 
		Next, we define the even and odd parity parts of the anomalous function as $f_{\mu,\pm}(E)=[f_\mu(x,x',E,\theta_S)\pm f_\mu(x',x,E,-\theta_S)]/2$, with $\mu\!=\!0,3$. 
		Once the anomalous function in \cref{eq:f-sym} is fully symmetric with respect to spin and parity, it belongs to one of four possible symmetry classes. 
		We label the different classes by their spin state, S for singlet and T for triplet, and their symmetry with respect to frequency and space coordinates, even (E) or odd (O). For example, odd-frequency, singlet, odd-parity is OSO and OTE refers to the odd-frequency, triplet even-parity case. 
		Consequently, the spin-singlet, even parity anomalous function reads
		\begin{gather}
		f_{0,+}(x,x',E,\theta_S)= \frac{i}{4\hbar} \left\{\begin{array}{lr}
		\begin{split}
		\e^{-i\frac{\epsilon}{\hbar}( \frac{x}{v_\up} + \frac{x'}{v_\dw} ) } 
		\left[ 
		\frac{a_1^\dw(\epsilon,\theta_S)}{v_\dw} \e^{i(k_\up x - k_\dw x')} 
		- 
		\frac{a_1^\up(\epsilon,-\theta_S)}{v_\up} \e^{-i(k_\up x - k_\dw x')} \right] 
		\\
		+\e^{-i\frac{\epsilon}{\hbar}( \frac{x}{v_\dw} + \frac{x'}{v_\up} ) } 
		\left[ 
		\frac{a_1^\dw(\epsilon,-\theta_S)}{v_\dw} \e^{-i(k_\dw x - k_\up x')} 
		- 
		\frac{a_1^\up(\epsilon,\theta_S)}{v_\up} \e^{i(k_\dw x - k_\up x')} \right] 
		\end{split}
		,& E>0 \\
		\begin{split}
		\e^{-i\frac{\epsilon}{\hbar}( \frac{x}{v_\up} + \frac{x'}{v_\dw} ) } 
		\left[ 
		\frac{a_1^\dw(\epsilon,\theta_S)}{v_\dw} \e^{i(k_\up x - k_\dw x')} 
		- 
		\frac{a_1^\up(\epsilon,-\theta_S)}{v_\up} \e^{-i(k_\up x - k_\dw x')} \right] 
		\\
		+\e^{-i\frac{\epsilon}{\hbar}( \frac{x}{v_\dw} + \frac{x'}{v_\up} ) } 
		\left[ 
		\frac{a_1^\dw(\epsilon,-\theta_S)}{v_\dw} \e^{-i(k_\dw x - k_\up x')} 
		- 
		\frac{a_1^\up(\epsilon,\theta_S)}{v_\up} \e^{i(k_\dw x - k_\up x')} \right] 
		\end{split}
		,& E<0
		\end{array}\right. ,
		\end{gather}
		which is the same for positive and negative energies. One can also check that $f_{0,+}(x,x';E,\theta_S)=f_{0,+}(x',x;E,-\theta_S)$, which indicates that $f_{0,+}(x,x';E,\theta_S)\equiv f_\text{ESE}(x,x';E,\theta_S)$. 
		
		We now focus on the local anomalous function with $x=x'\equiv x_0$. 
		We define the singlet and triplet Andreev scattering amplitudes as
		$a_{S,T}= [a_1^\up/v_\up \mp a_1^\dw/v_\dw ]/2$ and, using the short-hand notation $a^{\alpha}_{\beta}\!=\!(\alpha E, \beta \theta_S)$, with $\alpha,\beta=\pm$ and $a$ standing for either of $a_{S,T}$, we can define the energy- and angle-symmetric parts of the amplitude $a$ as
		\begin{subequations}\label{eq:EOPM2}
			\begin{align}
			a^{Ee}(E) ={}& \frac{1}{4} \left[ a^{+}_{+}(E) + a^{-}_{+}(E) + a^{+}_{-}(E) + a^{-}_{-}(E) \right] , \quad &
			a^{Eo}(E) ={}& \frac{1}{4} \left[ a^{+}_{+}(E) + a^{-}_{+}(E) - a^{+}_{-}(E) - a^{-}_{-}(E) \right] , \\
			a^{Oe}(E) ={}& \frac{1}{4} \left[ a^{+}_{+}(E) - a^{-}_{+}(E) + a^{+}_{-}(E) - a^{-}_{-}(E) \right] , \quad & 
			a^{Oo}(E) ={}& \frac{1}{4} \left[ a^{+}_{+}(E) - a^{-}_{+}(E) - a^{+}_{-}(E) + a^{-}_{-}(E) \right] . 
			\end{align}
		\end{subequations}
		Here, the index $a^{E,O}$ ($a^{e,o}$) labels the even and odd parts of the amplitude with respect to the energy (angle). 
		We can now apply the condition in \cref{eq:PHS-det-bal} to the singlet and triplet amplitudes and find that $a_S(E,\theta_S)=a_S^*(-E,-\theta_S)$ and $a_T(E,\theta_S)=-a_T^*(-E,-\theta_S)$, explicitly,
		\begin{equation}\label{eq:ST-complex}
		a_{S,T}(E,\theta_S)=\frac{a_\up(E,\theta_S)}{2v_\up} \mp \frac{a_\dw(E,\theta_S)}{2v_\dw} \overset{\text{[\cref{eq:PHS-det-bal}]}}{=} -\frac{a^*_\dw(-E,-\theta_S)}{2v_\dw} \pm \frac{a^*_\up(-E,-\theta_S)}{2v_\up} = \pm a^*_{S,T}(-E,-\theta_S)
		\end{equation}
		
		Expressing the singlet and triplet amplitudes in terms of the energy and angle-symmetric ones we find
		\begin{align}
		a_S(E,\theta_S)={}& a_S^{Ee} + a_S^{Oe} + a_S^{Eo} + a_S^{Oo} , \\
		a_S^*(-E,-\theta_S)={}& \left(a_S^{Ee}\right)^* - \left(a_S^{Oe}\right)^* - \left(a_S^{Eo}\right)^* + \left(a_S^{Oo}\right)^* , \\
		a_T(E,\theta_S)={}& a_T^{Ee} + a_T^{Oe} + a_T^{Eo} + a_T^{Oo} , \\
		-a_T^*(-E,-\theta_S)={}& - \left(a_T^{Ee}\right)^* + \left(a_T^{Oe}\right)^* + \left(a_T^{Eo}\right)^* - \left(a_T^{Oo}\right)^* ,
		\end{align}
		Then, using \cref{eq:ST-complex} and the fact that the symmetric amplitudes are linearly independent, we find that 
		\begin{align}\label{eq:complex-amps1}
		a_S^{Ee}= \left(a_S^{Ee}\right)^* , \quad 
		a_S^{Oo}=\left(a_S^{Oo}\right)^* , \quad
		a_T^{Oe}= \left(a_T^{Oe}\right)^* , \quad
		a_T^{Eo}= \left(a_T^{Eo}\right)^* ,
		\end{align}
		and these amplitudes are thus real. Analogously, we have that 
		\begin{align}\label{eq:complex-amps2}
		a_S^{Oe}=- \left(a_S^{Oe}\right)^* , \quad 
		a_S^{Eo}=- \left(a_S^{Eo}\right)^* , \quad
		a_T^{Ee}=- \left(a_T^{Ee}\right)^* , \quad
		a_T^{Oo}=- \left(a_T^{Oo}\right)^* ,
		\end{align}
		and these amplitudes are thus purely imaginary. 
		Therefore, when necessary, we can change the purely imaginary amplitudes as $a=i\bar{a}$, with $\bar{a}$ real. We show an example of the angle dependence of the fully-symmetric scattering amplitudes in \cref{fig:theta}. The region between the dashed vertical gray lines corresponds to $|\theta_S|\leq\theta_c$, where \cref{eq:complex-amps1,eq:complex-amps2} are valid. Using \cref{eq:EOPM2}, the local singlet components reduce to
		\begin{align}
		f_\text{ESE}(E)={}& \frac{-i}{2\hbar}\e^{-i\frac{\epsilon x_0}{\hbar v_\up v_\dw}( v_\up + v_\dw ) } 
		\left\{\begin{array}{lr}
		(a_S^{Ee} + a_S^{Oe})\cos[(k_\up-k_\dw)x_0] - i (a_T^{Ee} + a_T^{Oe})\sin[(k_\up-k_\dw)x_0] , & E>0 \\ 
		(a_S^{Ee} + a_S^{Oe})\cos[(k_\up-k_\dw)x_0] - i (a_T^{Ee} + a_T^{Oe})\sin[(k_\up-k_\dw)x_0] , & E<0 
		\end{array} \right. , 
		\\
		f_\text{OSO}(E)={}& \frac{-i}{2\hbar}\e^{-i\frac{\epsilon x_0}{\hbar v_\up v_\dw}( v_\up + v_\dw ) } 
		\left\{\begin{array}{lr}
		(a_S^{Eo} + a_S^{Oo})\cos[(k_\up-k_\dw)x_0] - i (a_T^{Eo} + a_T^{Oo})\sin[(k_\up-k_\dw)x_0] , & E>0 \\ 
		-(a_S^{Eo} + a_S^{Oo})\cos[(k_\up-k_\dw)x_0] + i (a_T^{Eo} + a_T^{Oo})\sin[(k_\up-k_\dw)x_0] , & E<0 
		\end{array} \right. ,
		\end{align}
		which fulfill the appropriate symmetries with respect to frequency and parity. Analogous expressions can be obtained for the triplet components. Interestingly, each symmetry class has a singlet-triplet mixing term proportional to $\sin[(k_\up-k_\dw)x_0]$, which is zero for any $x_0$ if $k_\up=k_\dw$, i.e., for \gls{nis} and \gls{nfis} junctions.
		
		\begin{figure}
			\includegraphics[width=0.90\textwidth]{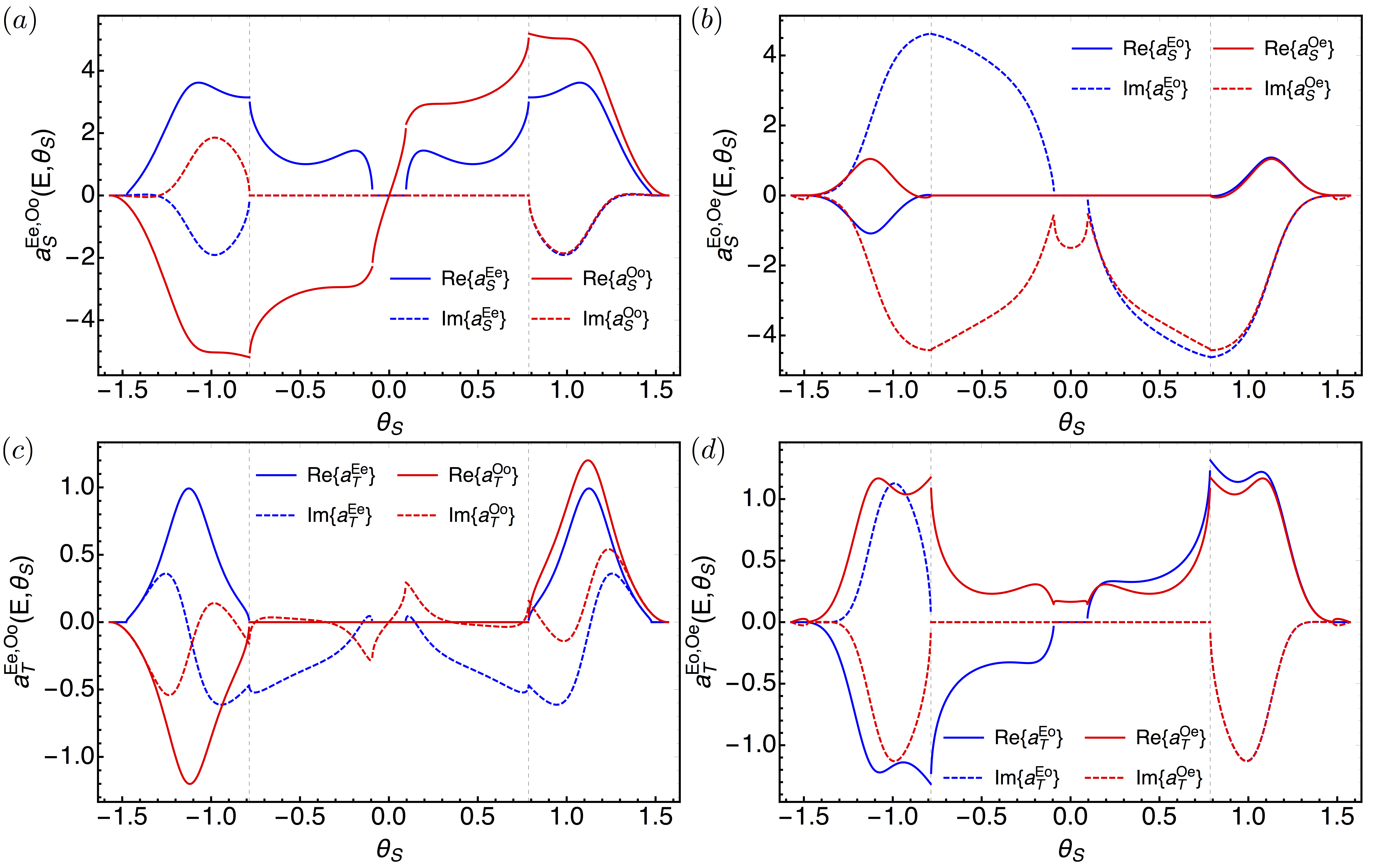}
			\caption{\label{fig:theta} 
				Fully symmetric singlet (a,b) and triplet (c,d) Andreev amplitudes as a function of the angle $\theta_S$. Real and imaginary parts are plotted with solid and dashed lines, respectively. The vertical gray dashed lines mark the value of $\theta_c$. 
				In all cases, the superconductor has two-dimensional $d_{xy}+is$-wave pairing, with $\Delta_1=0.80$, and the junction parameters are $X=0.5$, $T=0$, $Z_0=3$ and $Z_m=0$. 
			}
		\end{figure}
		
		At the interface with $x_0=0$ we find
		\begin{subequations}\label{eq:pair-int}
			\begin{align}
			f_\text{ESE}^\text{int} ={}& \left\{ \begin{array}{rclr}
			-\frac{i}{2\hbar}(a_S^{Ee}+a_S^{Oe}) &=& -\frac{i}{\hbar}[a_S(\epsilon,\theta_S)+a_S(\epsilon,-\theta_S)], & E>0 \\
			-\frac{i}{2\hbar}(a_S^{Ee}+a_S^{Oe}) &=& -\frac{i}{\hbar}[a_S(\epsilon,\theta_S)+a_S(\epsilon,-\theta_S)], & E<0 \end{array} \right. , 
			\\
			f_\text{OSO}^\text{int} ={}& \left\{ \begin{array}{rclr}
			-\frac{i}{2\hbar}(a_S^{Eo}+a_S^{Oo}) &=& -\frac{i}{\hbar}[a_S(\epsilon,\theta_S)-a_S(\epsilon,-\theta_S)], & E>0 \\
			+\frac{i}{2\hbar}(a_S^{Eo}+a_S^{Oo}) &=& +\frac{i}{\hbar}[a_S(\epsilon,\theta_S)-a_S(\epsilon,-\theta_S)], & E<0 \end{array} \right. , 
			\\
			f_\text{ETO}^\text{int} ={}& \left\{ \begin{array}{rclr}
			-\frac{i}{2\hbar}(a_T^{Eo}+a_T^{Oo}) &=& -\frac{i}{\hbar}[a_T(\epsilon,\theta_S)-a_T(\epsilon,-\theta_S)], & E>0 \\
			-\frac{i}{2\hbar}(a_T^{Eo}+a_T^{Oo}) &=& -\frac{i}{\hbar}[a_T(\epsilon,\theta_S)-a_T(\epsilon,-\theta_S)], & E<0 \end{array} \right. , 
			\\
			f_\text{OTE}^\text{int} ={}& \left\{ \begin{array}{rclr}
			-\frac{i}{2\hbar}(a_T^{Ee}+a_T^{Oe}) &=& -\frac{i}{\hbar}[a_T(\epsilon,\theta_S)+a_T(\epsilon,-\theta_S)], & E>0 \\
			+\frac{i}{2\hbar}(a_T^{Ee}+a_T^{Oe}) &=& +\frac{i}{\hbar}[a_T(\epsilon,\theta_S)+a_T(\epsilon,-\theta_S)], & E<0 \end{array} \right. .
			\end{align}
		\end{subequations}
		
		\section{Connection to the thermoelectric current}
		
		We now connect the previous results to the thermoelectric current. We focus on the Andreev contribution to \cref{eq:ch-current_sym}, which is equal to the total current in the cases discussed in the main text. First, the Andreev conductance is given by
		\begin{equation}\label{eq:cond_1}
		\sigma_\text{A}=R_\text{N} \mean{ P_\up \lambda_\up |a_{1\up}|^2 + P_\dw \lambda_\dw |a_{1\dw}|^2 }_{\mbf{k}_\parallel} 
		= R_\text{N} \mean{ v_\up v_\dw \left( |a_{S}|^2 + |a_{T}|^2 + 2 X \mathrm{Re} \left\{ a_S a_T^* \right\} \right)}_{\mbf{k}_\parallel} ,
		\end{equation}
		and the thermoelectric current is
		\begin{align}\label{eq:current_1}
		I_\text{A} ={}& \frac{1}{eR_\text{N}}\int_{-\infty}^{\infty}\! dE\, \delta f \sum\limits_{\sigma=\up,\dw}\mean{ P_\sigma \lambda_\sigma |a_1^\sigma(E,\theta_S)|^2}_{\mbf{k}_\parallel} = \frac{1}{eR_\text{N}}\int_{0}^{\infty}\! dE\, \delta f \sum\limits_{\sigma=\up,\dw}\mean{ P_\sigma \lambda_\sigma \left[|a_1^\sigma(E,\theta_S)|^2-|a_1^\sigma(-E,\theta_S)|^2\right]}_{\mbf{k}_\parallel} \notag \\
		={}& \frac{1}{eR_\text{N}}\int_{0}^{\infty}\! dE\, \delta f \mean{ 
			v_\up v_\dw \left[ |a_S(E,\theta_S)|^2-|a_S(-E,\theta_S)|^2 + |a_T(E,\theta_S)|^2-|a_T(-E,\theta_S)|^2 
			\right. \notag \\
			& \left.
			+ 2X\mathrm{Re}\left\{ a_S(E,\theta_S)a_T^*(E,\theta_S) - a_S(-E,\theta_S)a_T^*(-E,\theta_S) \right\} \right] 
		}_{\mbf{k}_\parallel} ,
		\end{align}
		where we have used that $a_1^{\up,\dw}=v_{\up,\dw}(a_T\pm a_S)$ from the definition of the singlet and triplet components, $\lambda_{\sigma}=v_{\bar{\sigma}}/v_\sigma$, and $P_{\up(\dw)}=[1+(-)X]/2$. 
		It is clear in \cref{eq:cond_1,eq:current_1} that a finite polarization ($X\neq 0$) couples the spin singlet and triplet states, resulting in a new contribution to the current. 
		
		We can now identify the scattering amplitudes with the anomalous Green's function described above. 
		First, we note that the virtual Andreev processes do not contribute to the current and thus the angle average is limited to $|\theta_S|\leq\theta_c$. 
		Using \cref{eq:EOPM2,eq:complex-amps1,eq:complex-amps2} we can write the scattering amplitudes in the absence of virtual processes as follows:
		\begin{subequations}\label{eq:f-amp-conection}
			\begin{align}
			a_S(E)={}& a_S^{Ee} + a_S^{Oe} + a_S^{Eo} + a_S^{Oo} = (a_S^{Ee} + i \bar{a}_S^{Oe}) + (a_S^{Oo} + i \bar{a}_S^{Eo}) \equiv 2i\hbar f_\text{ESE}^\text{int} + 2i\hbar f_\text{OSO}^\text{int} , 
			\\
			a_S(-E)={}& a_S^{Ee} - a_S^{Oe} + a_S^{Eo} - a_S^{Oo} = (a_S^{Ee} - i \bar{a}_S^{Oe}) - (a_S^{Oo} - i \bar{a}_S^{Eo}) \equiv 2i\hbar \left(f_\text{ESE}^\text{int}\right)^* - 2i\hbar \left(f_\text{OSO}^\text{int}\right)^* , 
			\\
			a_T(E)={}& a_T^{Ee} + a_T^{Oe} + a_T^{Eo} + a_T^{Oo} = (a_T^{Oe} + i \bar{a}_T^{Ee}) + (a_T^{Eo} + i \bar{a}_T^{Oo}) \equiv 2i\hbar f_\text{OTE}^\text{int} + 2i\hbar f_\text{ETO}^\text{int} , 
			\\
			a_T(-E)={}& a_T^{Ee} - a_T^{Oe} + a_T^{Eo} - a_T^{Oo} = -(a_T^{Oe} - i \bar{a}_T^{Ee}) + (a_T^{Eo} - i \bar{a}_T^{Oo}) \equiv -2i\hbar \left(f_\text{OTE}^\text{int}\right)^* +2i\hbar  \left(f_\text{ETO}^\text{int}\right)^* , 
			\end{align}	
		\end{subequations}
		where we have used that the amplitudes $a_S^{Ee}$, $a_S^{Oo}$, $a_T^{Eo}$ and $a_T^{Oe}$ are real, while the rest are purely imaginary and thus can be recast as $a=i\bar{a}$, cf.~\cref{eq:complex-amps1,eq:complex-amps2}. 
		
		Inserting \cref{eq:f-amp-conection} into \cref{eq:cond_1,eq:current_1}, the Andreev conductance reads
		\begin{equation}
		\sigma_\text{A}=R_\text{N} \mean{ v_\up v_\dw \left( |f_\text{ESE}+f_\text{OSO}|^2 + |f_\text{ETO}+f_\text{OTE}|^2 + 2 X \mathrm{Re} \left\{ (f_\text{ESE}+f_\text{OSO}) \left(f_\text{ETO}+f_\text{OTE}\right)^* \right\} \right)}_{\mbf{k}_\parallel} ,
		\end{equation}
		and the current becomes
		\begin{equation}
		I_\text{A} = 
		\frac{16\hbar^2}{eR_\text{N}} \int_{0}^{\infty}\! dE\, \delta f \mathrm{Re} 
		\left\{ \mean{ v_\up v_\dw \left(
			\left(f_\text{ESE}^\text{int}\right)^* f_\text{OSO}^\text{int} + \left(f_\text{ETO}^\text{int}\right)^* f_\text{OTE}^\text{int} +
			X \left[ \left(f_\text{ESE}^\text{int}\right)^* f_\text{OTE}^\text{int} + \left(f_\text{ETO}^\text{int}\right)^* f_\text{OSO}^\text{int} \right] \right) }_{\mbf{k}_\parallel} \right\} .
		\end{equation}
		
		The angle-average of the product of functions of opposite parity in the same spin state should be zero, i.e., $\mean{f_\text{ESE}^*f_\text{OSO}}_{\mbf{k}_\parallel}\!=\!\mean{f_\text{ETO}^*f_\text{OTE}}_{\mbf{k}_\parallel}\!=\!0$, unless the rotational symmetry is broken by $\alpha\!\neq\!0$. That is not the case for the products of different spin states, $\mean{f_\text{ESE}^*f_\text{OTE}}_{\theta_S}$ and $\mean{f_\text{ETO}^*f_\text{OSO}}_{\theta_S}$, which then result in the thermoelectric current for \gls{fis} junctions even when $\alpha\!=\!0$. 
		
	\end{widetext}

%

%
%

\end{document}